\documentclass[%
reprint,
nofootinbib,
 amsmath,amssymb,
prd,
]{revtex4-1}

 \pdfoutput=1

\usepackage{hyperref}
\usepackage{graphicx}
\usepackage{dcolumn}
\usepackage{bm}

\usepackage{xcolor}

\newcommand{\iMpc}{\,h\mathrm{Mpc}^{-1}}

\def\bea{\begin{eqnarray}}
\def\eea{\end{eqnarray}}
\def\be{\begin{equation}}
\def\ee{\end{equation}}
\def\fr{\frac}

\def\ci{\cite}
\def\la{\label}
\def\be{\begin{equation}}
\def\ee{\end{equation}}

\def\le{\left}
\def\ri{\right}
\def\Lc{\Lambda_c}

\def\L{\Lambda}
\def\ac{a_c}
\def\LCDM{$\Lambda$CDM\,}

\def\ad{a_{\star}}

\def\rex{\rho_{ex} }
\def\Oex{\Omega_{ex} }

\def\Hsmx{H_{smx} }
\def\Hsm{H_{sm} }

\begin{document}

\title{Cosmological signatures of a Rapid Diluted Energy Density} 

\author{Axel de la Macorra}
\email{macorra@fisica.unam.mx}
\affiliation{Instituto de F\'isica, Universidad Nacional Aut\'onoma de M\'exico, Cd. de M\'exico C.P. 04510, M\'exico} 
\affiliation{Instituto de Ciencias del Cosmos, University of Barcelona ICCUB, Barcelona 08028 Spain }
 
\author{Dante V. Gomez-Navarro} 
\email{dgomezn@estudiantes.fisica.unam.mx}
\affiliation{Instituto de F\'isica, Universidad Nacional Aut\'onoma de M\'exico, Cd. de M\'exico C.P. 04510, M\'exico}

\author{Alejandro Aviles}
\email{avilescervantes@gmail.com}
\affiliation{Consejo Nacional de Ciencia y Tecnolog\'ia, Av. Insurgentes Sur 1582, Colonia Cr\'edito
Constructor, Del. Benito Ju\'arez, 03940, Ciudad de M\'exico, M\'exico,} 
\affiliation{Departamento de F\'isica, Instituto Nacional de Investigaciones Nucleares,
Apartado Postal 18-1027, Col. Escand\'on, Ciudad de M\'exico,11801, M\'exico}

\author{Mariana Jaber}
\email{jaber@astro.umk.pl}
\affiliation{Institute of Astronomy, Faculty of Physics, Astronomy and Informatics,  Nicolaus Copernicus University, Grudziadzka 5,  87-100 Toru\'n, Poland}

\author{Jorge Mastache}
\email{jhmastache@mctp.mx}
\affiliation{Mesoamerican Centre for Theoretical Physics, Universidad Aut\'{o}noma de Chiapas,  Carretera Zapata Km. 4, Real del Bosque, 29040, Tuxtla Guti\'{e}rrez, Chiapas, M\'{e}xico,}
\affiliation{Consejo Nacional de Ciencia y Tecnolog\'ia, Av. Insurgentes Sur 1582, Colonia Cr\'edito
Constructor, Del. Benito Ju\'arez, 03940, Ciudad de M\'exico, M\'exico}

\author{Erick Almaraz}
\email{almaraz@fisica.unam.mx}
\affiliation{African Institute for Mathematical Sciences AIMS, Cape Town, South Africa}

\date{\today}

\begin{abstract}
We study the cosmological signatures of having extra energy density, $\rho_{ex}$, beyond the $\Lambda$CDM model that dilutes rapidly, faster than radiation, at a  scale factor $a_c$ with a corresponding mode $k_c=a_c H(a_c)$ crossing the horizon at that time. These types of models are motivated by phase transitions of the underlying elementary particles, for example the creation of protons and neutrons from almost massless quarks or the recently proposed Bound Dark Energy model. The rapidly dilution of $\rho_{ex}$ leaves distinctive imprints in the Universe not only in the expansion history with a clear impact on the acoustic scale, $r_s(a_cc)$, and angular distances, $D_A(a)$, but also in the matter and CMB power spectra. The rapidly diluted energy density  $\rho_{ex}$, (RDED) generates characteristic signatures that can be observed with current and future precision cosmological data. In particular, we find a bump in the matter power spectrum compared to the standard $\Lambda$CDM. We identify the amplitude, width, and time scale of the bump to the physical properties of the transition. We study these effects with linear theory, standard perturbation theory, and the correlated impact on cosmological distances, allowing for independent measurements of these extensions of the standard  $\Lambda$CDM model.

\end{abstract}

\maketitle


\section{Introduction}\label{introduction}

In the last two decades the amount of precision cosmological data have confirmed 
the discovery of the accelerated expansion rate of the Universe \cite{Riess:1998cb, Perlmutter:1998np}  and has improved significantly our understanding of the Universe by probing different scales and regimes with increasing precision. 
Data from the temperature and polarization of the Cosmic Microwave Background (CMB) radiation \cite{Aghanim:2018eyx}, galaxy and quasar surveys \cite{Alam:2020sor,Ahumada:2019vht, Aghamousa:2016zmz, 2009arXiv0912.0201, laureijs2011euclid}, lensing probes \cite{de2013kilo} or supernovae catalogs \cite{Scolnic:2017caz}, have consolidated our picture of a flat $\Lambda$CDM model undergoing recently a phase of accelerated expansion. 
However, despite its success, $\Lambda$CDM  suffers theoretical issues that have motivated alternative models challenging its assumptions. Most of these extensions look for plausible explanations of the dark sector, both dark matter and dark energy,
which together accounts for about 96\% of the energy content of the Universe at present time.

The increasing amount of clustering data and the percentage precision for cosmic distances \cite{Alam:2020sor,Aghanim:2018eyx} allows to search for extensions beyond $\Lambda$CDM by looking for cosmological features in for example the matter or CMB power spectra,  standard distances rulers, or tensions in \LCDM model as the recent $H_0$ crisis \cite{Riess:2019cxk}.
The increasing statistical tension in the estimated Hubble parameter from early and late times observations \cite{Verde:2019ivm} has reignited interest in alternative models.  Some of these models add extra components to the energy-momentum tensor that modify the expansion rate of the Universe at different epochs \citep{Poulin:2018cxd,Keeley:2019esp} or to understand the origin of dark energy by introducing extra particles within a gauge group \cite{Almaraz:2018fhb,delaMacorra:2018zbk}, similar as in the standard model of particle physics, leaving potentially observable signatures. For the interest of this work, we consider phase transition in the dark sector generating localized features, from now on ``bumps'', in the matter power spectrum for modes entering the horizon around the phase transition time.

In this work, we study the cosmological signatures of having an extra density component in the Universe, $\rho_{ex}(a)$ that rapidly dilutes, faster than radiation components, at a scale factor $a_c$. We refer to such component as Rapid Diluted Energy Density (RDED). It may appear in different cosmological models, for example in the recently proposed Bound Dark Energy model (BDE)   \citep{Almaraz:2018fhb,delaMacorra:2018zbk}, where the original elementary particles (e.g quarks) form neutral massive bound states, like protons and neutrons in the strong QCD force of the standard model, and the lightest scalar field corresponds to dark energy. Alternatively, we can study the dynamics of the dark sector in a model-independent way \citep{Bassett:2002qu, Corasaniti:2002vg}, or in dark energy models having a steep equation of state (SEOS) \cite{Jaber:2017bpx}. 

The RDED transition affects the background evolution, due to the change in the total energy density, consequently modifying the cosmological distances and the evolution of matter fluctuations leaving imprints in the matter and CMB spectra \citep{Pogosian:2005ez,Almaraz:2019zxy, Jaber:2019opg}. 
Linear density perturbations are enhanced for modes crossing the horizon just before the RDED transition takes place. That is, for modes $k \geq k_c \equiv \ac H(\ac)$, the increase in $H$, compared to $\Lambda$CDM, is reflected in a higher growth rate of matter perturbations in the radiation-dominated era creating a  distinctive feature in the matter power spectra  a bump, more visible when we take the ratio  $P_{\text{RDED}}(k)/P_{\text{$\Lambda$CDM}}(k)$. 
Its amplitude becomes related to the amount of extra energy density $\rex$ that dilutes, while its width, to the duration of this transition (i.e., how many modes are affected). Modes with $k < k_c$  do not share this boost since the amount of energy density and background evolution are the same as in $\Lambda$CDM. A similar bump is also generated when the RDED
transition happens at a later time in a matter or dark energy dominated universe.

We will consider  BDE and SEOS like models as our motivation to further study the cosmological features of these types of phase transition models.  We implement these models with the Boltzmann code \textsc{camb}\footnote{\href{http://camb.info/}{http://camb.info/}} \cite{Lewis:1999bs}, varying the abundance of the RDED and duration of the transition to generating bumps with different shapes. We then follow up on these features through  1-loop SPT computations.

In BDE, the bumps are primarily located at nonlinear scales, out of reach of perturbation theory; however, their effects start to be important at large, yet linear scales, hence a treatment within SPT is reliable over a small interval of its full range, and we generally observe an enhancement of their amplitudes and shift of their peaks toward slightly larger scales. On the other hand, the transition for the model SEOS occurs at very late times and henceforth the signatures are located at very large scales and are not affected by non-linearities.

This paper is organized as follows: in Sec.~\ref{sec_2} we present the imprints to the Hubble parameter and the cosmological distances due to the extra energy density. In Sec.~\ref{sec.P} we show how the evolution of energy density perturbations change due to the RDED and how it is reflected in the matter power spectrum $P(k)$. We present our results and details for the different models of study in Sec.~\ref{results}. Finally, the conclusions are in Sec.~\ref{sec.conclusion}. In appendix \ref{appendix} we include some scalar field models energy  that could render the RDED transition.

 
\section{Rapidly Diluted Energy Density (RDED) }\label{sec_2}

In this section, we study the cosmological  imprints  and consequences  of having an extra energy density  $\rex(a)$ beyond the standard $\Lambda$CDM  that dilutes rapidly at a scale factor $a_c$. 
 Such a dilution may be generated by a phase transition of the underlying particle model or due to the dynamical properties of the equation of state of the fluid, for example  for quintessence  dark energy models.  In the latter case, the RDED component can be described in terms of a scalar field $\phi$,  with evolution depending on the choice of a potential $V(\phi)$ and kinetic energy. 
 Here, we will focus on the implications of this energy component over the background expansion history and on cosmological distances.

The RDED transition has been suggested in the BDE model, by means of a phase transition occurring at early times, well inside the radiation dominated epoch  $a_c\sim 10^{-6}$, affecting modes $k\sim 1 \iMpc$ that are entering the horizon at those times. At a much later time, close to present day, a second rapid dilution takes place in BDE  \cite{delaMacorra:2018zbk},
where the EoS goes from $w_i=-1$ for $z\gg 1$ to  $w_0(z=0)=-0.93$ at present time, with an intermediate value $w(z_c) =-0.965$ at a redshift  $z_c = 0.625$, due to the dynamics of the dark energy scalar field. We see that BDE contains two epochs encountering a RDED component one at early times and the second close to the present time.

Alternatively, the late time behaviour of DE has also been investigated in a model independent analysis  by introducing a phenomenological EoS $w(a)$, modeling a phase transition with an abrupt change at $a_c$. In this SEOS models \cite{Jaber:2017bpx, Jaber:2019opg} we found a late time transition at $z_c=0.28$  with $w(z\gg 1)=-1$ and $w_0=-0.93$, with a lower value of $\chi^2_{BAO+H_0}$ compared to $\Lambda$CDM, and consistent with the BDE model. Both BDE and SEOS adjust very well to observations; yet, some  Bayesian criteria suggest a better fitting to the data than $\Lambda$CDM \cite{delaMacorra:2018zbk,Jaber:2017bpx}.

BDE has well-defined parameter values, with little room for adjustments, and hence the generated bumps have precise locations and widths. However, in this work, we generalize BDE in order to extract generic properties of having a RDED component regardless of the method that generates it. Nevertheless, we will follow the guideline that BDE generates bumps at small scales since the phase transition occurs at early times, and we will refer as SEOS bumps to those generated at large scales since the transitions occur at late times; referring to those mechanisms simply as BDE and SEOS.

In the rest of this section and in the next section, we will study the implications of having  $\rex$ over cosmological distances and linear fluctuations of matter densities.

\subsection{Analytic Model Independent Analysis}\label{analytic}

We present a model independent framework  of RDED. 
We consider the standard model of cosmology and add  an extra energy density $\rex$, that is present only at times before the transition, $a \leq a_c$.

The Hubble parameter $H\equiv\dot a/a$   is determined by the Friedman equation
\begin{equation} \label{Hsm} 
H^2_{sm}(a) = \tfrac{8\pi G}{3}\,\rho_{sm},
\end{equation}
with
\begin{equation}
   \rho_{sm} =  \rho_{mo}a^{-3}+ \rho_{ro}a^{-4}+ \rho_{\Lambda}
\end{equation}
the standard model total energy density. The matter density $\rho_m$  contains DM and baryons,  $\rho_{r}$ are the relativistic particles, i.e. photons and neutrinos which we consider massless, while $\rho_{\Lambda}$ corresponds to dark energy. The subindex $o$ denotes present time quantities.
For scale factors $a\leq a_c$, the extra energy density  $\rex (a)$ contributes to the total energy density, and Friedmann equation becomes  
\be
H^2_{smx}(a) = \tfrac{8\pi G}{3}\,\rho_{smx} = \tfrac{8\pi G}{3}\,\le(\rho_{sm}+\rex \ri).
 \label{Hsmx} 
\end{equation}
Notice that we have used the subscript ``$sm$'' to refer to the standard components,  while the subscript ``$smx$'' corresponds to adding  $\rho_{ex}$. The amount of extra energy density is given by
\be
\Oex \equiv \fr{\rex}{\rho_{sm}+\rex} = 1 - \fr{H^2_{sm}(a)}{H^2_{smx}(a)},
\la{oex}\ee
which in the radiation dominated epoch can be approximated by 
\be
\Omega_{ex} \simeq \fr{N_{ex}\,\beta}{1+(N_\nu+N_{ex})\,\beta},
\la{Oex}\ee
with 
$\rex=\beta N_{ex} T_\nu^4$, $T_\nu$ the neutrinos temperature,
$N_\nu=3.046$ for three massless neutrinos and $\beta \equiv \le(7/8\ri)  \le(4/11\ri)^{4/3}$.

To gain physical insight, we first consider the simplified case on which $\rex$ tracks the dominant background energy density, such that $w_{ex} \simeq w_{sm}$ and $\Oex$ becomes a constant. Later on, in  section \ref{results}, we will use \textsc{camb} to follow the exact evolution of $\Oex$.
A RDED takes place if the EoS  $w_{ex}\equiv p_{ex}/\rex$  suffers  a transition from  $w_c\equiv w_{ex}(a \leq \ac)=w_{sm}$  to  $w_f \equiv w(a_f > \ac)> w_c$.
To have a RDED the value of  $\Delta w\equiv  w_f  -w_c>0$  must be positive, while
the width $\Delta a\equiv  (a_f - a_c)/a_c$  sets the steepness of  the transition 
and how fast $\rex$ dilutes compared to $\rho_{sm}$ for $a>\ac$, with
\be \label{rho_ex_rho_sm}
\frac{\rex}{\rho_{sm}} \propto \le(\fr{a}{a_c}\ri)^{- \Delta w}.
\ee 
A slow transition impacts more wave-number modes $k$ than a fast one.
Notice that a particle with mass $m$ goes from being relativistic, at early times with $T/m \gg 1 $, and hence $w=w_c=1/3$,  to  non-relativistic at late times, such that $w_f=0$. Hence, one expects  $\Delta w = w_f-w_c = -1/3 $ to be negative. The interesting cases in RDED happen when $\Delta w$ is positive, and hence this transition is beyond the standard model. 
However, such phase transitions can be realized in dynamical scalar fields, where one finds examples where the EoS goes from $w_c=1/3$ or $w_c=0$ to $w_f=1$ at radiation or matter domination epochs, or from $w_c\simeq -1$ to $w=-0.93$ close to the present time  \cite{Almaraz:2018fhb,delaMacorra:2018zbk, Jaber:2017bpx,delaMacorra:1999ff,Steinhardt:1999nw,Copeland:2006wr}.

In  Appendix A  we show examples of scalar fields yielding indeed this type of transition in the EoS.

A transition due to a RDED taking place at $\ac$ modifies $\Hsm$, and henceforth  the comoving angular distance $D_A$, the angular diameter $D_M=c/H$, the acoustic scale at recombination $r_s$ and the diffusion damping scale $r_d$. 
All these distances are well constrained  by CMB \cite{Aghanim:2018eyx}, BAO \cite{Anderson2014, Alam:2016hwk, Alam:2020sor}  and SNIa  \cite{Scolnic:2017caz} observations  allowing us to constrain the cosmological  models.
Besides the changes in cosmological distances, the evolution of perturbations will be affected and 
we determine the matter and CMB power spectra in section \ref{results}.

\subsection{Impact on Cosmological Distances}\la{sec.D}

Let us now study how an extra energy density $\rex$, for $a\leq \ac$ and diluting at $\ac$,  affects cosmological distances.
The precise value of the distances requires  to be numerically calculated, however approximated
analytic expressions  of the cosmological distances give us an simple  understanding on how
this $\rex$  modifies them, mainly due to a change in the Hubble parameter $H$.

The sound horizon and  damping scale at recombination $\ad$ are given by
\be
r_s(\ad)=\int^{\ad}_{0}  \fr{c_s\,da}{a^2H(a)},
\la{rs}\ee
with $c_s$ to the sound speed of the photon-baryon plasma,
\be \label{cs}
c_s(a)=\fr{1}{\sqrt{3(1+R)} },
\ee
with $R = (3/4)(\rho_b/\rho_\gamma)$  carrying the speed of sound time dependence. An approximate expression for the damping scale is given by \cite{Menegoni:2012tq}
\be
r_d^2 (\ad)= (2\pi)^2 \int_{0}^{\ad} \fr{  da }{ a^3\sigma_T n_e H} \le[ \fr{R^2 +\frac{16}{15}(1+R)}{6(1+R^2)}\ri],
\la{rdif}\ee
with  $n_e$  the number density of free electrons,  $\sigma_T $   the Thomson scattering cross-section, and the factor in between square brackets is due to the directional and polarization dependence of the electron-photon scattering. 

A non-vanishing $\rex$ in the region $a\leq \ad$ will 
affect the value of the Hubble parameter $H$, modifying 
$r_s(\ad)$ and $r_d (\ad)$, and impacting the CMB  and BAO observations, with the caveat that for BAO measurements one has to consider the 
drag scale  $a_{drag} \simeq 9.4 \times 10^{-4}$  instead of the recombination scale $\ad \simeq 1/1090$ as the upper integration limit. 
Cosmological distances such as the angular diameter distance at recombination
\be 
D_A(\ad)=\int^{a_o}_{\ad} \fr{da}{a^2H(a)}, 
\la{dan}\ee
will not be affected if $\ac<\ad$. On the other hand, 
if the transition occurs after recombination ($\ac>\ad$), then the acoustic scale, the  damping  scale and  in general cosmological distances, such as 
\be
d_L(a)=\fr{1}{a} \int_{a}^{a_o} \fr{da'}{a'^2 H(a')},\;\;\;\;  D_M(a)=\fr{c}{H(a)} ,
\la{dm}\ee
relevant for  SN1a and BAO measurements are modified  by a non-vanishing $\rex$ in $H(a)$ in eqs.(\ref{dan}) and (\ref{dm}).

\subsection{$\Oex$ constant}

Let us present  the analytic solution assuming $\Oex$ constant for $a \leq \ac$,
valid if $\rex$ tracks the leading contribution on $H$, and $\Oex=0$ for $a>\ac$. In this limit we can express
\be
\fr{\Hsm}{\Hsmx} =\sqrt{1-\Oex}
\label{HH}
\ee
which is constant.  
Since all the cosmological distances considered in the previous subsection the integrand is proportional to $1/H$
we can then simply  obtain the ratio  of  $r_s(a)$  for $a\leq \ac$ in the two models as
\bea
r_s^{smx}(\ac) &\equiv &  \int^{\ac}_{0} \fr{c_s\, da}{a^2H_{smx}}
=\int^{\ac}_{0} \le(\fr{H_{sm}}{H_{smx}}\ri)\fr{c_s\, da}{a^2H_{sm}} \nonumber\\
 &=& \sqrt{1 -\Oex} \int^{\ac}_{0} \fr{c_s\, da}{a^2H_{sm}},
\eea
and express it in terms of $r_s^{sm}(\ac)$ to obtain
\be
\frac{r_s^{smx}(\ac)}{r_s^{sm}(\ac)} = \sqrt{1 -\Oex },
\label{rsoex}\ee
such that the acoustic scale at the transition scale is suppressed by the presence of the extra density component.
Similar equations hold generally for cosmological distances,
\be
D^{smx} (\ac)= \sqrt{1-\Oex}\; D^{sm} (\ac).
\la{ddr} 
\ee
Of course, these results  are valid as long as $\Oex$ remains constant  which  requires $\rex$ to  track (i.e. to have the same equation of state)  as the leading energy density in $H_{smx}$, which is a reasonable working hypothesis. If $\rex$  dilutes faster than the background it will have little or no effect, meanwhile if $\rex$ dominates $H$ it will be ruled out by observations. 

\subsubsection{Region with $ a_c < \ad$ }

Here we will study the model when $a_c < \ad$. In this case  we can distinguish two scenarios having  $a_c < a_{eq}$ or
$ a_{eq}< a_c <\ad$. In the first  case let us consider for simplicity and presentation purposes that the Universe is dominated by radiation and we consider $c_s$ constant. 
The quantity  $\Delta r_s(\ad) \equiv  r_s^{smx}(\ad)- r_s^{sm} (\ad)$ can be divided in two integrals
from $a_i$ to $\ac$ and from $\ac$ to $\ad$. The second integration will cancel since both term in $\Delta r_s(\ad)$ have the same integrand for $a_c\leq a\leq \ad$ and we than have
\be
\Delta r_s (\ad) =   \Delta r_s (\ac)  
\la{rsrc}\ee
where the r.h.s of Eq.~(\ref{rsrc}) is evaluated from $a_i\leq a\leq a_c$,
with $\Delta r_s (\ac) =  r_s^{smx}(\ac) -r_s^{smx}(\ac)$. Using  Eq.~(\ref{rsoex}) we simply get
\be
\fr{\Delta r_s (\ad)}{r_s^{sm}(\ad)}=\fr{\Delta r_s (\ac)}{r_s^{sm}(\ad)}=  \le(\sqrt{1-\Oex} -1\ri) \le(\fr{r_s^{sm} (\ac)}{r_s^{sm}(\ad)}\ri).
\ee
We see  that $\Delta r_s(\ad)\neq 0$  due to the contribution of  $\Oex$.  
On the other hand  for  $a_c < \ad$  the angular diameter distance $D_A(\ad)$ is not affected by $\Oex$ and we have 
\be
\Delta D_A (\ad)= D_A^{smx}(\ad) - D_A^{sm}(\ad)=0.
\ee

\subsubsection{ Region with $\ad < a_c  $}

We consider now a transition for  $ \ac > \ad$  with $\Oex$ constant. The ratio of acoustic scale  $r_s$  is
\be
 \fr{r_s^{smx} (\ad)}{ r_s^{sm}(\ad)}= \sqrt{1-\Oex},
\ee
while  for the angular distance we get
\bea
D_A^{smx}(\ad) &=&\int^{a_c}_{\ad} \fr{da}{a^2 H_{smx} }+\int^{a_o}_{a_c} \fr{da}{a^2H_{sm}}  \\
&=& \sqrt{1-\Oex}  \int^{a_c}_{\ad} \fr{da}{a^2H_{sm}} + \int^{a_o}_{a_c} \fr{da}{a^2H_{sm}}. \nonumber 
\eea
We compare $D_A^{smx}(\ad)$  with  $D_A^{sm}(\ad)$ by  taking their difference,  $\Delta D_A(\ad)  \equiv D_A^{smx}(\ad)  - D_A^{sm}(\ad)$, yielding
\bea
\Delta D_A (\ad) &=& \le(\sqrt{1-\Oex} -1\ri) \int^{a_c}_{\ad} \fr{da}{a^2H_{sm}} \nonumber\\
&=& \le(\sqrt{1-\Oex} -1\ri) D_A(\ac)
\la{DDA} \eea
which is negative for $\Oex \neq 0$; and therefore, we see that $\Oex$ reduces $D_A^{smx}(\ad)$ compared to $D_A^{sm}(\ad)$.


\section{Signatures of a Rapid Dilution  of $\rho_{ex}$  on  Density Perturbation}\label{sec.P}

We will now study the impact  of  a RDED model on structure growth and the signals it leaves in the matter power spectrum. 
We will show that modes entering the horizon before the RDED transition takes place at $a_c$ will grow faster for a non-vanishing $\rex$. This growth generates a bump in the linear matter power spectrum, easily noticeable by taking the quotient of the spectra of models with and without the RDED component. The bump is located for wave-vectors with $k\geq k_c$,  corresponding to  modes entering the horizon at a scale factor $a\leq a_c$ with an associated mode
\be
k_c\equiv \ac H(a_c).
\la{kc}\ee

\subsection{Linear density evolution }

Let us now consider the evolution of matter densities $\delta_m$ in our two models; that we will refer  for notational simplicity, as  $X$ for the $smx$ model, and  $\Lambda$ for the standard model $sm$. The effect of having extra particles $\rex$ impacts  the amplitude of the energy density perturbation $\delta_m(a_h)=\delta \rho_m(a_h)/\rho_m(a_h)$, the scale factor at horizon crossing ($a_h$) for the same mode $k$ (i.e. $k=a_h^XH(a_h^X)=a_h^\L H(a_h^\L)$) and the subsequent evolution.

\subsubsection{Outside Horizon}\label{out}

Outside the horizon the amplitude of the density modes $k$ remain constant, but once they enter the horizon they start to grow. 
The ratio at horizon crossing  is \cite{Ma:1995ey}
\be
\fr{\delta_{m}^X(a_h^X)}{\delta_{m}^{\Lambda }(a_h^\L)}=\fr{1+\frac{4}{15} f_\nu^{\Lambda }}{1+\frac{4}{15}f_\nu^{X}},
\la{dxdl}\ee
where
\begin{align}
f_\nu^\Lambda= \frac{\rho_{\textrm{eff}}^\Lambda}{\rho_\gamma+\rho_{\textrm{eff}}^\Lambda}, \qquad&  \rho_\textrm{eff}^\L=\rho_\nu, \\
f_\nu^X= \frac{\rho_{\textrm{eff}}^X}{\rho_\gamma+\rho_{\textrm{eff}}^X}, \qquad&   \rho_\textrm{eff}^X=\rho_\nu+\rho_{ex}.
\end{align}
account for the relativistic particles contribution. Since $f_\nu^\Lambda/f_\nu^X \leq 1$ one has 
a lower amplitude for $smx$ model $\delta_{m}^X(a_h^X)\leq \delta_{m}^{\Lambda}(a_h^\Lambda)$.

A fixed mode $k$ that crosses the horizon in the $\Lambda$ model at the scale factor $a_h^\L$, given by  $k=a_h^\L H^\L(a_h^\L)$,  would otherwise enter the horizon in the $X$ model at a scale factor $a_h^X$, given by $k=a_h^XH^X(a_h^X)$. Therefore, we get
\be
\fr{a_{h}^X}{a_{h}^{\Lambda}} =\fr{H^\L(a_h^\L)}{H^X(a_h^X)}=\sqrt{\fr{1-f_\nu^{\L}}{1-f_\nu^X}} =\fr{1}{\sqrt{1-\Oex}}.
\label{aa}\ee
Notice that the presence of an extra $\rex$  component provokes a mode $k$ to enter the horizon  at a later time   $a_{h}^X > a_{h}^{\Lambda}$  
with  $H^\L(a_h^\L)> H^X(a_h^X)$.

\subsubsection{Linear density evolution at early times}

At horizon crossing modes with $k > k_c$ in $X$   and $\L$ models  have a relative scale factor at horizon crossing given by  eq.(\ref{aa}) 
$a^X_h/a^\L_h =1/{\sqrt{1-\Oex}} > 1$.  Modes  in the $X$ model cross the horizon a later time and have therefore  less  time to grow and it is reflected as an early suppression in  $\Delta \delta_m$. 
However, after the initial suppression at horizon crossing the matter perturbations in the $X$  model  have a higher growing rate than in the $\Lambda$CDM model that not only compensates but also reverses the initial suppression.

To gain physical intuition on how the rapid diluted dark energy  component affects matter density fluctuations well inside the radiation dominated epoch we analyze a simplified version of the equations (ultimately all these quantities will be computed using the code \textsc{camb}). To this end, let us solve $ \delta_m''+\mathcal{H} \delta_m'=0 $ where a prime means derivative with respect to conformal time $\tau$ and $\mathcal{H} =aH$ is the conformal Hubble rate.
The solution   is 
\be
\delta_m'(\tau)= \fr{\tau_i}{\tau}= \fr{aH}{a_i H_i}
\la{deltatau}\ee
with  initial condition $1/\tau_i=\mathcal{H}_i =a_i H_i$ at some pivotal scale $a_i$, which can be taken as $a_c$.
We can see that for $a< a_c$ we have an increase rate for $\delta_m$ for $\rex >0$ and it
ceases for modes entering the horizon for $a>a_c$. Therefore the  increase takes place only for modes $k\geq k_c $, while modes with smaller $k$  enter  the horizon after the RDED dilution has taken place and both models  have the same expansion rate and  evolution of matter perturbations. 

Alternatively,  we find convenient to present the evolution in terms of the scale factor,  the  solution  for $\delta_m$ is simply a logarithmic function ---the Mezsaros effect---  $ \delta_m(a)  =   \delta_{mi}  + ( \delta_{mi}'/k) \;\textrm{ln}\le ( a/a_h \ri)$ with  initial conditions   $\delta^{'}_{m i} /k_i=2\delta_{m i}$, giving
\bea
\delta_m^X (a)  &=& \delta_{mi}^X  \le(\textrm{ln}(a/a_h^X)+1/2 \ri), \\
\la{dmL}\delta_m^\L (a)  &=& \delta_{mi}^\L  \le(\textrm{ln}(a/a_h^\L)+1/2 \ri).
\eea

This enhancement can be semi analytically estimated in terms of the dilution of the extra component $\rex$  after $a_c$. For simplicity we assume that the dilution takes place at $a_c$ and therefore $H$ contains the extra relativistic particles (we name it $H_+^X$) for $a\leq a_c$ while  $H_-^X$ for $a>a_c$ does no longer have them. This change is reflected in the value of  initial condition  $\delta_m'= \tau_i/\tau$   where we can take the initial condition
at  $a_i= a_c$ with  $\tau_{i-} =1/\mathcal{H}_-^X=1/(a_cH_-^X)$,  $\tau_{i+} =1/\mathcal{H}_+^X=1/(a_cH_+^X)$  and a ratio
$\tau_{i-}/\tau_{i+} = H_+^X / H_-^X$.
We  obtain  then
\be
\delta_{m}^X(a) = \delta_{mi}^X \le[ \textrm{ln} ( a_c/a_h^X) + (H_+^X/H^X_-) \textrm{ln} (a/a_c)+1/2\ri]  
\la{dmX}\ee
for $a>a_c$  with
\be
\fr{H_+^X}{H_-^X}=\fr{1}{\sqrt{1-\Oex} }.
\la{hh}\ee
To easily compare Eq.~(\ref{dmX}) to Eq.~(\ref{dmL}) we write $a_c/a_h^X=(a_c/a_h^\L)(a_h^\L/a_h^X)$
\be
\delta_{m}^X(a) = \delta_{mi}^X \le[ \le(\fr{H_+^X}{H^X_-}\ri) \textrm{ln} \le(\fr{a}{a_c}\ri) + \textrm{ln} \le( \fr{a_c}{a_h^\L} \fr{ a_h^\L}{a_h^X} \ri) +\fr{1}{2} \ri]  
\ee
and leads to a  ratio $\Delta \delta_m=\delta_m^X/\delta_m^\L$
\be
\Delta \delta_m =  \fr{ \delta_{mi}^X}{\delta_{mi}^\L}
 \fr {\le[  \le(\fr{H_+^X}{H^X_-}\ri) \textrm{ln} \le(\fr{a}{a_c}\ri) +\textrm{ln} \le( \fr{ a_h^\L}{a_h^X}\ri) + \textrm{ln} \le( \fr{a_c}{a_h^\L}\ri)+\fr{1}{2}   \ri] }
{  \textrm{ln}\le(\fr{a}{a_c}\ri) + \textrm{ln}\le(\fr{a_c}{a_h^\L}\ri) +\fr{1}{2} } ,
\la{DD}\ee
with  $a_c/a_h^\L= k/k_c$  ($k=a_h^\L H^\L (a_h^\L)$) and $a_h^\L/a_h^X$, $H_+^X/H^X_-$ given in  Eqs.~(\ref{aa})  and (\ref{hh}), respectively. Equation (\ref{DD}) is  valid for $a>a_c$  and $ k>k_c$.  
From  Eq.~(\ref{DD}) we see an increase of $\Delta \delta_m$   which tends to 
$\delta_{mi}^X/\delta_{mi}^\L$ for  $k\gg k_c$, while the enhancement is $\delta_m^X/\delta_m^\L-1=  (\delta_{mi}^X/\delta_{mi}^\L)(H_+^X/H_-^X)-1$ 
for  $a\gg \ac$. Since $H_+^X/H_-^X > 1$   we see that the linear matter perturbation $\delta_m$ in radiation domination grows faster for $\rex>0$. 

The bump is generated because the linear growth  $\delta_m(k)$ has a higher increase rate 
in radiation domination for larger $H$, as seen from $\delta_m'= \tau_ i/\tau=\mathcal{H}/\mathcal{H}_i\propto aH $. 
This increase is only valid  for modes $k\geq k_c $ while for mode $k<k_c$  both models have the same  $H$  and the evolution of $\delta_m(k)$ is the same in both cases. This explains why a  RDED model shows a bump
compared to $\Lambda$CDM. The amplitude of the bump  is related to the amount of $\rex$ while 
the width and steepness of the bump  are determined  by how fast the RDED transition takes place and can be
parametrized  phenomenological by the  quantities $\Delta a=(a_f - a_c)/a_c$ and  $\Delta w=w_f-w_c$.
The modes $k<k_c$ that enter the horizon after $a_c$ are not affected by the extra energy density  $\rex$  dilution,
for these modes we have  $H_+^X=H_-^X$ and $a^X_h=a^\Lambda_h$.
The final shape of the matter power spectrum  is a combination of the present value of $\delta_m$ determined by the dynamical processes described in this section and the fitting values of $n_s$ and $A_s$ which define the primordial spectrum $P_s$. Here we consider the same $P_s$ in $sm$ and  $smx$ models.

\subsubsection{Linear Matter density evolution in matter and dark energy  domination epoch}

In matter or Dark Energy domination the evolution of matter density perturbations are well described by  the ansatz
\be
f\equiv \fr{d \log \delta_m}{d \log a} =\Omega_m^{\gamma} (a)
\la{f}\ee
where  $\gamma\simeq 0.55$ in a $\Lambda$CDM model and the content of matter is given by $\rho_m^\L=\rho_m$
and $ \rho_m^X=\rho_m+\rex$.   
Let us consider the regime  with $\Omega_m$ constant. In this case the solution to Eq.~(\ref{f}) is just 
\bea
\delta_m^X (a)  &=& \delta_{mi}^X \le(\fr{a}{a_h^X}\ri)^{(\Omega_m^X)^\gamma} \la{dm1}\\
\delta_m^\L (a)  &=& \delta_{mi}^\L \le(\fr{a}{a_h^\L}\ri)^{(\Omega_m^\L)^\gamma}. 
\la{dm2}\eea
Well within  matter domination we can  set  $\Omega_m \simeq 1$ and  the ratio of $\delta_m$ in two models $X$ and $\Lambda$CDM is simply
\be
\Delta \delta_m  = \fr{\delta_m^X}{\delta_m^\L} = \le( \fr{\delta_{mi}^X}{\delta_{mi}^\L} \ri)  \le( \fr{a_h^\L}{a_h^X} \ri).
\ee
Comparing $\delta_m$ for the same mode $k$,  with $k=a_h^XH_c^X=a_h^\L H_c^\L$,  we get  from  Eq.~(\ref{HH})
\be
\fr{a_h^X}{a_h^\L}= \fr{H_c^\L}{H_c^X}=  \sqrt{1 -\Oex }
\ee
and taking the same initial condition at horizon crossing    $\delta_{mi}^X= \delta_{mi}^\L $  we arrive at
\be
\Delta \delta_m  = \fr{\delta_m^X}{\delta_m^\L} =  \fr{a_h^\L}{a_h^X} =  \fr{1}{ \sqrt{1 -\Oex }}.
\la{ddm}\ee
In the case where $\Omega_m$ is not equal one, we can solve  Eq.~(\ref{f})  by taking the  ratio   $f^q=(a/\delta^q_m)(d\delta^q_m/da) $ for  $q=X,\L$, i.e. 
\be
\fr{f^X}{f^\L}=\le(\fr{\delta_m^\L}{\delta_m^X}\ri)\le( \fr{d\delta_m^X}{d\delta_m^\L} \ri) =\le(\fr{\Omega_m^X}{\Omega_m^\L}\ri)^\gamma
\la{fxfl}\ee
and considering 
\be
\fr{\Omega_m^X}{\Omega_m^\L}=\le(\fr{\rho_m+\rho_{ex}}{\rho_m} \ri)\fr{(H^\L)^2}{(H^X)^2}=\le(1+ \fr{\rho_{ex}}{\rho_m} \ri) \le(1-\Oex \ri)
\la{omom}\ee
constant the solution to Eq.~(\ref{fxfl})  is 
\be
\log\le[\fr{\delta_m^X(a)}{\delta_{mi}^X}\ri]=\le( \fr{\Omega_m^X}{\Omega_m^\L}\ri)^\gamma  \log\le[ \fr{\delta_m^\L(a)}{\delta_{mi}^\L}\ri].
\la{dxdy}\ee
We see  in eqs.(\ref{dm1}), (\ref{dm2})   and  Eq.~(\ref{dxdy}) that  the linear perturbation $\delta_m$  in the regime of matter/dark energy domination grows faster  in the case with  $\rex >0$.


\section{Numerical Results}\la{results}

\begin{figure*}[ht]
        \includegraphics[height = 0.37\textheight]{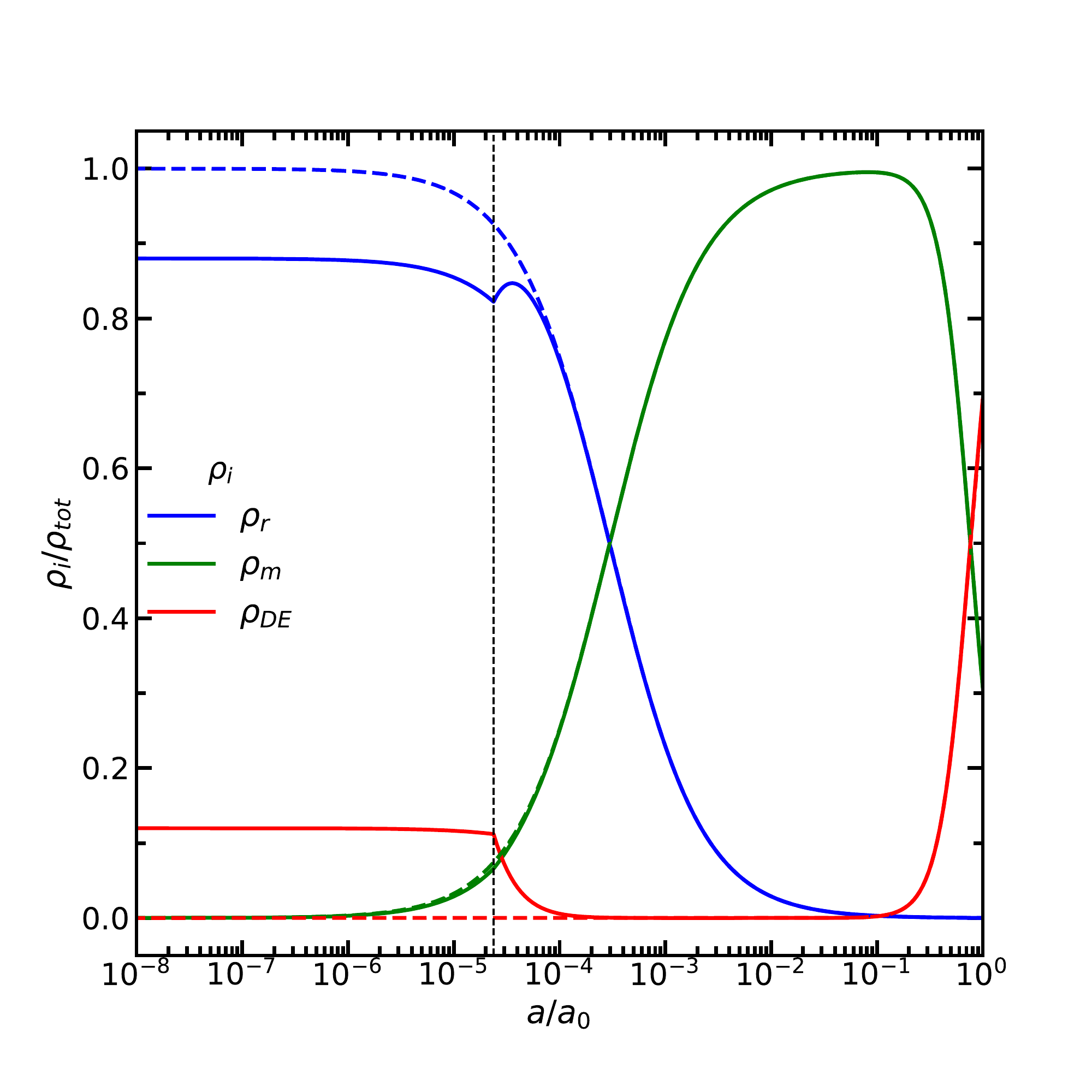}
        \includegraphics[height = 0.37\textheight]{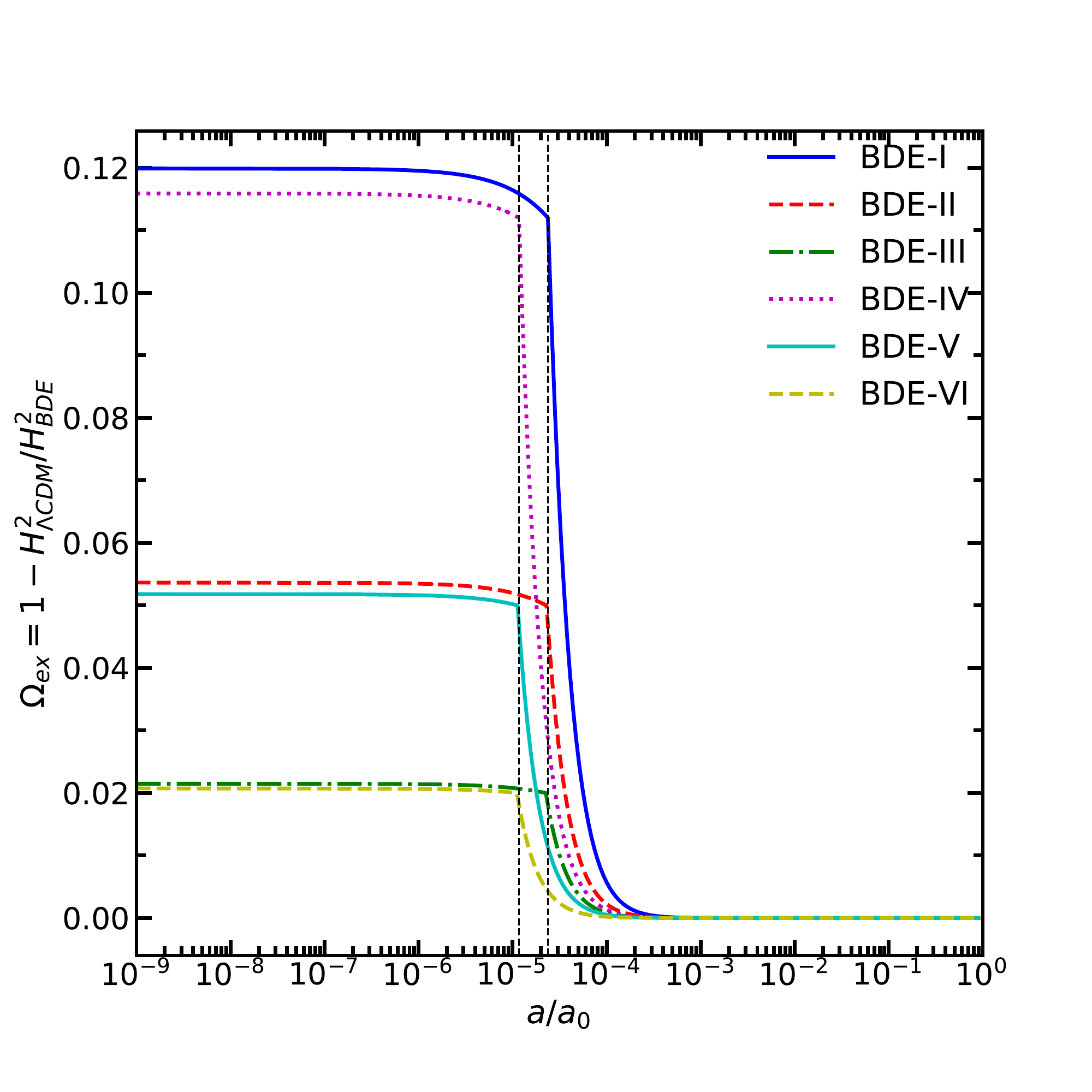}
\caption{\textbf{Evolution of densities and their corresponding amount of extra energy density.} Evolution of densities of different components for BDE (solid curves) and the standard $\Lambda$CDM model (dashed curves). The left panel shows the dynamical dark energy density (red curve) of BDE-I model with a dilution at $a_c = 2.37 \times 10^{-5}$ when it contributes $\approx 11 \%$ to the total density. Its contribution quickly decreases after that. The right panel shows the amount of extra energy density for BDE models described in Table \ref{tab:bde}.  Solid blue are for BDE-I; dashed red are for BDE-II; dash-dotted green are for BDE-III; dotted magenta are for BDE-IV; solid cyan are for BDE-V; and dashed yellow are for BDE-VI. The vertical lines represent the transition redshift for each model. We note that in the BDE models we have different transitions for each model (see Table \ref{tab:bde}).}
\label{fig:omega_omegaex_bde}
\end{figure*}
%

\begin{table*}
	\centering
	\begin{tabular}{lccccccc} 
		\hline
		& & & & & & & \\[-1pt]
		Model \;\;\; &  $k_c$ \;\;\; & $a_c[10^{-5}]$\;\; & $\Omega_{ex}^c$ \;\;& $k_{b,L}$ \;\;\; & $R_\text{L}(k_{b})$ \;\;\; &  $k_\text{b,1-loop}$ \;\;\; & $R_\text{1-loop}(k_{b,1-loop})$ \\[4pt]
		\hline
		\hline
		& & & & & & & \\[1pt]
		\textsc{bde-i} & $0.148$ & $2.373$ & $0.112$ & $0.474$ & $1.160$ & $0.917$ & $1.272$\\[4pt]
		\textsc{bde-ii} & $0.148$ & $2.292$ & $0.050$  & $0.474$ & $1.066$ & $0.917$ & $1.110$\\[4pt]
		\textsc{bde-iii} & $0.148$ & $2.255$ & $0.020$ & $0.474$ & $1.025$ & $0.917$& $1.041$ \\[4pt]
		\textsc{bde-iv} & $0.295$ & $1.168$ & $0.112$ & $1.061$ & $1.195$ & $1.908$ & $1.339$ \\[4pt]
		\textsc{bde-v} & $0.295$ & $1.129$ & $0.050$ & $1.061$ & $1.081$ & $1.773$ & $1.134$\\[4pt]
		\textsc{bde-vi} & $0.295$ & $1.111$ & $0.020$ & $1.061$ & $1.031$ & $1.773$ & $1.051$\\[4pt]
		\hline
	\end{tabular}
\caption{Small Scales - BDE cosmologies. We show  $k_c=a_cH(a_c)$ 
(in hMpc$^{-1}$) and   $a_c$ the mode and scale factor at the transition,  $\Omega_{ex}^c\equiv \Omega_{ex}(a_c)$,   $k_b$ and  $k_\text{b,1-loop}$ the modes at the maximum of the bump at a linear 
and one-loop level with $R_\text{L}(k_b)$ and $R_\text{1-loop}(k_b, \text{1-loop})$ the ratio of the power spectra 
($R(k)\equiv P_\text{BDE}(k)/P_\text{$\Lambda$CDM}(k)$) at a linear  and one-loop level, respectively.
We consider $h=0.677$, $\Omega_m=0.307$ for all bumps and a late time $\Lambda$CDM cosmology.}	
\label{tab:bde}
\end{table*}

We use a modified version of the code \textsc{camb}  and produce the linear matter spectra and the different multipoles of the matter power spectrum and CMB power spectrum for the BDE and SEOS cosmologies.

\subsection{Cosmological signatures}

\subsubsection{Transitions at early times}

In this section we are motivated by BDE and we focus on phase transitions occurring at early times, during the radiation-dominated era. In such a case the mode affected have wave-vectors $k > k_{eq}$.   In table \ref{tab:bde} we show the models we analyse, which are specified by a choice of  $k_c\equiv a_c H(a_c)$, $\Omega_{ex}^c\equiv \Omega_{ex}(a_c)$.

The left panel of Figure \ref{fig:omega_omegaex_bde} shows the evolution of energy densities of the different components as a function of the scale factor.
In red we have the contribution of extra relativistic energy density $\rex (a)$. We clearly notice the transition at $a_c =2.37\times 10^{-5}$ and the rapid dilution of $\Omega_{ex}$  for $a>a_c$. In the right panel we  plot the six different BDE models given  in Table \ref{tab:bde}).  We took two different modes $k_c$ with three different abundances $\Oex$ in each case. The value of $k_c$ and the amount of $\Oex$ and $k_c$ determines the value of $a_c$. We notice that $\Oex$ is constant for $a<a_c$ and  after  $a_c$ a steep phase transition takes place  with $\rho_{ex}$ redshifting  as $\rho \propto a^{-6}$  and disappears, e.g.   
by factor of $\Oex\propto (a/a_c)^{-2} \sim  1/100$  at $a/a_c=10$ in all the models.

On the other hand, the extra components for different BDE cosmologies are shown in the right panel. We consider six BDE models which differ from each other by the dilution epoch ($a_c$) and the amount of extra components at the transition ($\Omega_{ex}^c$). For instance, BDE-I and BDE-IV have the same amount of extra relativistic particles, but dilution occurs at different epoch $2.37 \times 10^{-5}$ and $1.17\times 10^{-5}$, respectively. Similarly for the BDE-II and BDE V models (both with $\Omega_{ex}^c=0.05$) but different $a_c=2.92\times10^{-5}$ and $a_c=1.13 \times 10^{-5}$, respectively; and for the BDE-III and BDE-VI models with same $\Omega_{ex}^c=0.020$ and different transition time, $a_c=2.25\times10^{-5}$ and $1.11\times 10^{-5}$, respectively. We note that the amount of extra relativistic particles is negligible at late time reproducing a cosmological constant behaviour. In section \ref{cmb-ps}  we present the effects of the extra relativistic particles on the CMB and matter power spectra.

\begin{figure*}[ht]
        \includegraphics[height = 0.37\textheight]{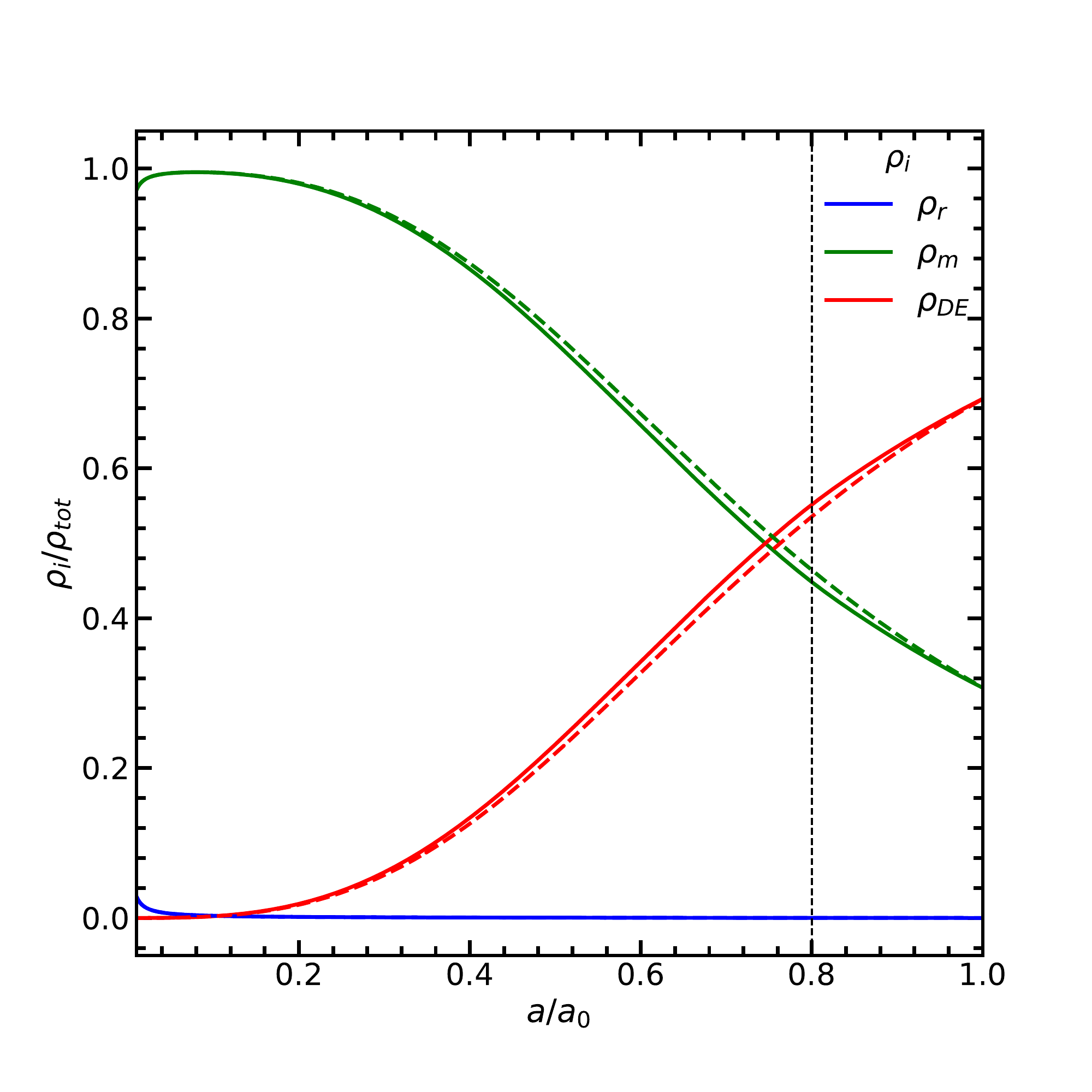}
        \includegraphics[height = 0.37\textheight]{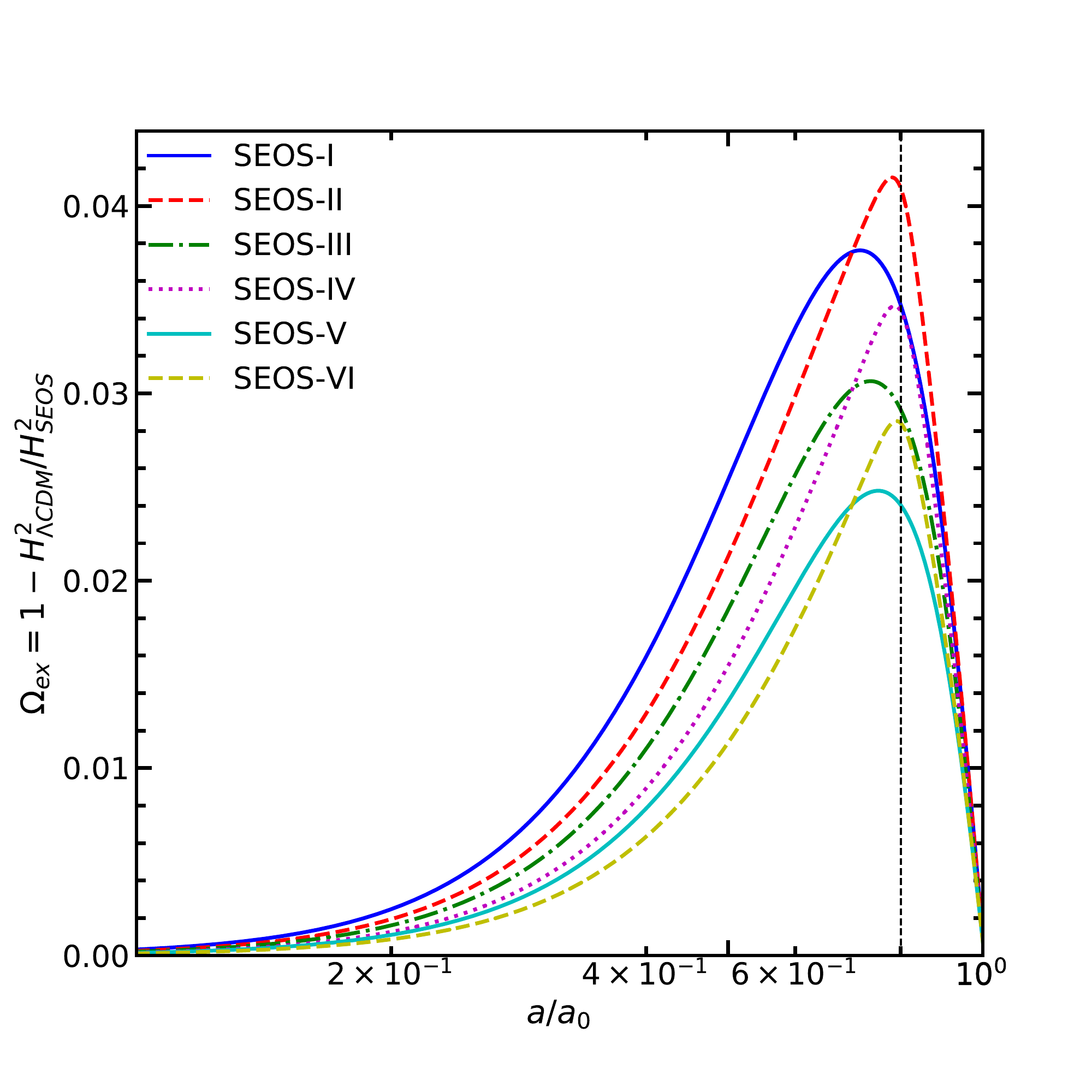}
\caption{\textbf{Evolution of densities and their corresponding amount of extra energy density $\rex$.} Left panel: Evolution of densities of different components for SEOS-I model (solid curves) and the standard $\Lambda$CDM model (dashed curves) with a dilution at $a_c = 0.8$ ($z_c=0.25$). Blue curves are for radiation density; green curves are for matter density; and red curves are for dark energy density. The right panel shows the amount of extra energy density for SEOS models described in Table \ref{tab:seos}. Solid blue are for SEOS-I; dashed red are for SEOS-II; dash-dotted green are for SEOS-III; dotted magenta are for SEOS-IV; solid cyan are for SEOS-V; and dashed yellow are for SEOS-VI. The vertical lines represent the transition scale factor $a_c=1/(1+z_c)=0.8$.}
\label{fig:omega_omegaex_seos}
\end{figure*}
%

\begin{table*}
	\centering
	\begin{tabular}{lccllllccccc} 
		\hline
		& & & & & & & & & & & \\[-1pt]
		Model &\;\;\;\; $w_0$ \;\;\; & \;\;\;\;\; $w_i$ \;\;\;\;\;\;& $z_c$\;\;\;\;\;\; & $q$ \;\;\;\;\;\; & $\Omega_{ex}^c(a_c)$ \;\; &  $k_c \times 10^{4}$  &\;\; h \;\;\; &\;\;  $\Omega_{de}(a_0)$\;\;\;\; & $\Omega_m (a_0)$\;\;\;\; &  $k_{b,L}\times 10^{4}$\;& $R_\text{L}(k_{b,L})$ \\[4pt]
		\hline
		\hline
		& & & & & & & & & & \\[-2pt]
		\textsc{seos-i} & $-0.9$ & $-1$ & $0.25$ & $2$ & $0.035$ & 3.19 & $0.80$ & 0.78  & $0.22$  & $4.95$ & $0.981$ \\[4pt]
		\textsc{seos-ii} & $-0.9$ & $-1$ & $0.25$ & $10$ & $0.041$ & 3.21 &$0.80$ & 0.78  & $0.22$ & $5.15$ & $0.982$ \\[4pt]
		\textsc{$\Lambda$cdm-i} & $-1$ & $-1$ & $-$ & $-$ & $-$ &   $-$  &$0.80$ & 0.78  & $0.22$ & $-$& $-$ \\[4pt]
		\textsc{seos-iii} & $-0.9$ & $-1$ & $0.25$ & $2$ & $0.029$ & 3.25  &$0.68$ & 0.69 & $0.31$ & $5.58$ & $0.985$ \\[4pt]
		\textsc{seos-iv} & $-0.9$ & $-1$ & $0.25$ & $10$ & $0.034$ & 3.27 & $0.68$ & 0.69 &$0.31$ &$6.05$ & $0.986$\\[4pt]
		\textsc{$\Lambda$cdm-ii} & $-1$ & $-1$ & $-$ & $-$ &  $-$ &  $-$ & $0.68$ & 0.69 & $0.31$ & $-$ & $-$ \\[4pt]
		\textsc{seos-v} & $-0.9$ & $-1$ & $0.25$ & $2$ & $0.024$ & 3.32  & $0.60$ & 0.61& $0.39$ & $6.05$ & $0.988$\\[4pt]
		\textsc{seos-vi} & $-0.9$ & $-1$ & $0.25$ & $10$ & $0.028$ & 3.33 & $0.60$ & 0.61& $0.39$& $6.05$ & $0.989$ \\[4pt]
		\textsc{$\Lambda$cdm-iii} & $-1$ & $-1$ & $-$ & $-$ & $-$ &  $-$ &  $0.60$ & 0.61&  $0.39$ &$-$ & $-$\\[4pt]
		\hline
	\end{tabular}
	\caption{
	Large Scales - SEOS cosmology. We present the values of the EoS parameters	$w_0$, $w_i$ $z_c$, $q$, 
	with $\Omega^c_{ex}\equiv \Omega_{ex}(a_c)$ for a transition mode $k_c \equiv a_c H(a_c)$ hMpc$^{-1}$ at $a_c =0.8$.  The amount of matter $\Omega_m h^2= 0.141$ is the same in all cases.
	The mode at the maximum of the bump is $k_b$ in [$\iMpc$] units while
	 $R_\text{L}(k_{b,L}) \equiv P_\text{SEOS}(k_{b,L})/P_\text{$\Lambda$CDM}(k_{b,L})$) is the ratio of the linear power spectra.}
\label{tab:seos}
\end{table*}

\begin{figure*}[ht]
        \includegraphics[height = 0.37\textheight]{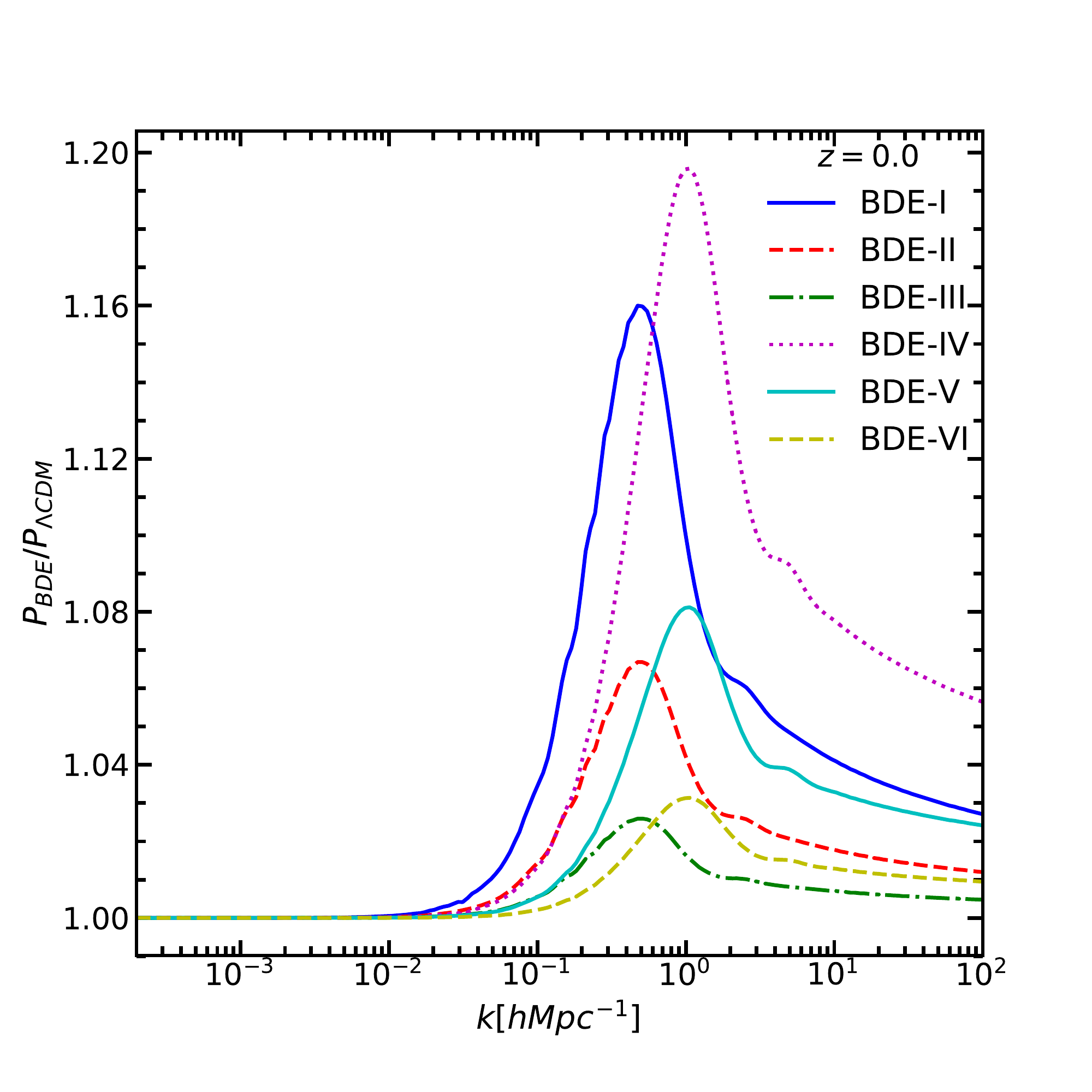}
\caption{\textbf{Linear matter power spectrum for BDE models.} The ratio $P_{BDE}/P_{\Lambda CDM}$ for different BDE models at $z=0$. Solid blue are for BDE-I; dashed red are for BDE-II; dash-dotted green are for BDE-III; dotted magenta are for BDE-IV; solid cyan are for BDE-V; and dashed yellow are for BDE-VI.}
\label{fig:power-linear-bde}
\end{figure*}
\begin{figure*}[ht]
        \includegraphics[height = 0.37\textheight]{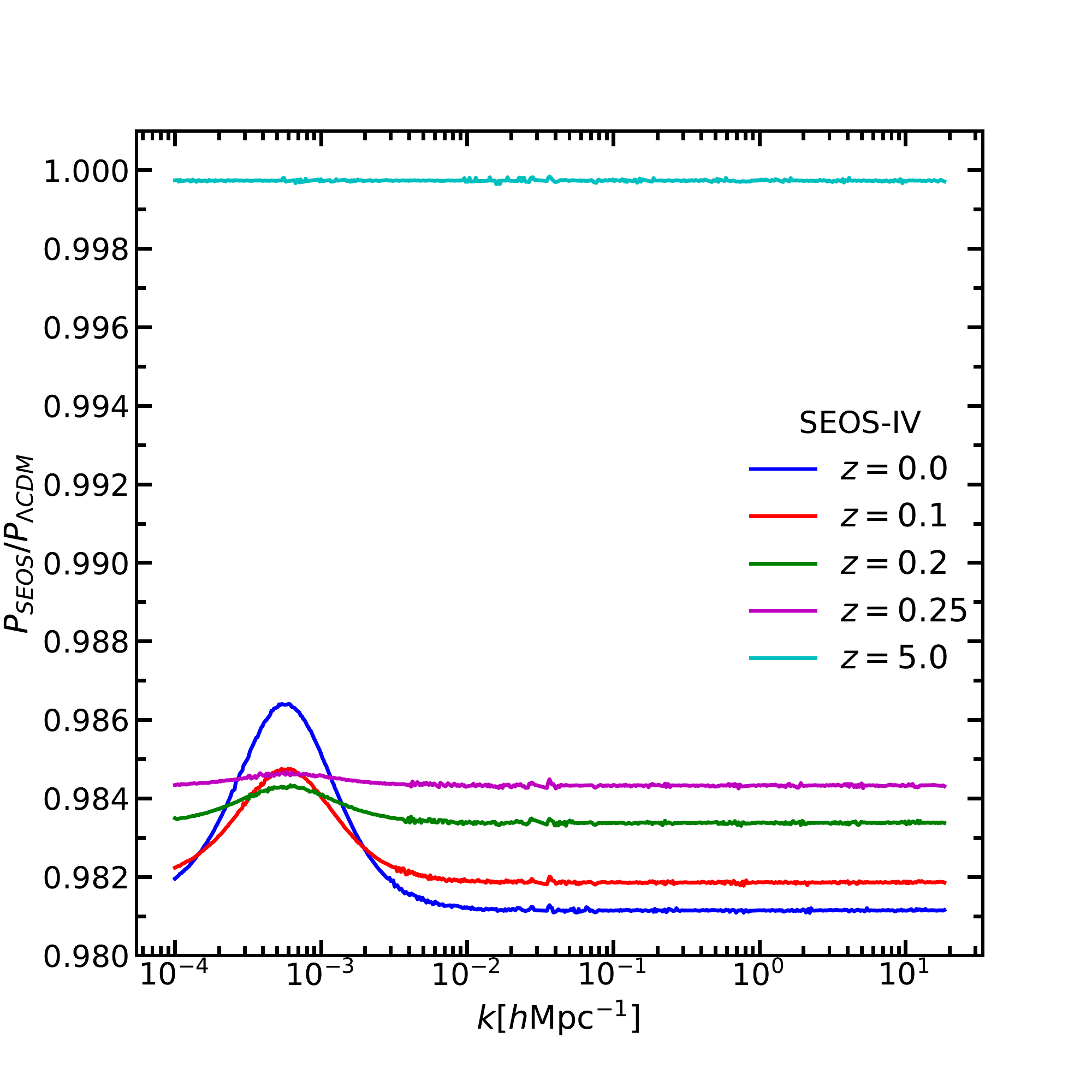}
        \includegraphics[height = 0.37\textheight]{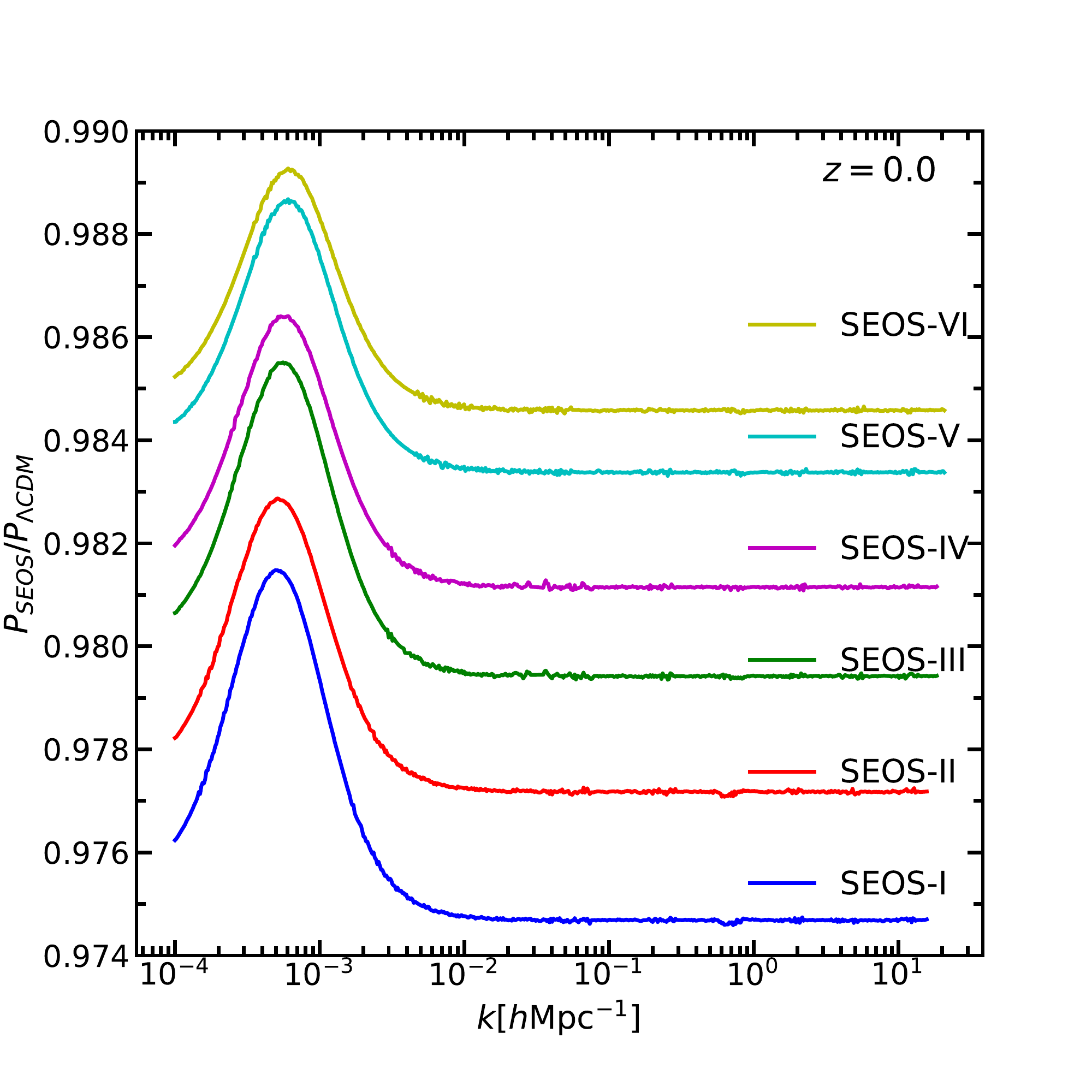}
\caption{\textbf{Linear matter power spectrum for SEOS models.} The left panel:  the ratio $P_\text{SEOS}/P_\text{$\Lambda$CDM}$for model \textsc{SEOS-IV}. Blue curves are for $z=0.0$; red curves are for $z=0.1$; green curves are for $z=0.2$; magenta curves are for $z=0.25$; and cyan curves are for $z=5.0$. The right panel: The ratio $P_\text{SEOS}/P_\text{$\Lambda$CDM}$ for different models at $z=0$. Blue curves are for SEOS-I; red curves are for SEOS-II; green curves are for SEOS-III; magenta curves are for SEOS-IV; cyan curves are for SEOS-V; and yellow curves are for SEOS-VI (see Table\ref{tab:seos}).}
\label{fig:power-linear-seos}
\end{figure*}

 \subsubsection{Transitions at late times}
 
Here we present the case where the transition takes place at late times  and the dark energy  is modelled as a barotropic fluid, where its dynamics is parameterized in terms of an Steep Equation of State  (``SEOS" \cite{Jaber:2017bpx}), $w(z)$: 
\be
\label{wseos}
 w(z) = w_0 + (w_i-w_0)\,\le(\frac{(z/z_c)^q}{1+(z/z_c)^q}\ri),
\ee
where $w_i$, $w_0$, $z_c$ and $q$ are free parameters, with $z_c>0$ and a finite value of $q$. 
In this case we  have a RDED transition occurring at late times given by the redshift $z_c$ and are due to the evolution of the dynamical the dark energy model. The evolution of Dark Energy evolves from $w_i$ for $z\gg z_c$ and has a transition from $w_i$ to $w_0=w(z=0)$ 
at a redshift $z_c$ with an EoS $w(z_c)=(w_i-w_0)/2$.
A best fit value $w_0=-0.93$ and was obtained in  in SEOS  \cite{Jaber:2017bpx}  and BDE  \cite{delaMacorra:2018zbk}  models. 

A cosmological constant $w\equiv -1$  can be recovered using Eq.~(\ref{wseos}) setting  $w_0=w_i=-1$ for all $z$ and  independent of the values of $z_t$, and $q$.
Left panel  in Fig.~\ref{fig:omega_omegaex_seos} shows the evolution of the energy  densities of different components, including the corresponding amount of dark energy at different redshifts for the SEOS model. The evolution of  $\Omega_i=\rho_i/\rho_{tot}$ for SEOS-I model with $q =2$, $\Omega_{de}(a_0) = 0.78$, $h = 0.80$ and for $\Lambda$CDM, are shown in the left panel.

The EoS in Eq.~(\ref{wseos}) allows for a steep transition from $w_i$ to $w_0$  taking place at a central redshift value  $z_c$ with a steepness determined  by the parameter $q$. The effect of the dynamical dark energy in SEOS model is seen in the right panel of Fig.~\ref{fig:omega_omegaex_seos} where the parameter $q$ modulates the steepness of the transition, a larger $q$ has
a steeper transition.  We show in table \ref{tab:seos}  and in Fig.~\ref{fig:omega_omegaex_seos} 
different  SEOS models where we take as examples the values of $q=2$ and $q=10$ and  we allow for different amount of $\Omega_{de}(a_0)$ at present time. 
A larger value of $q$ has a steeper transition, originating a narrower bump for  $q=10$ in models  SEOS-II, SEOS-IV, and SEOS-VI models,  while  broader bump is generated  for $q=2$ in SEOS-I, SEOS-III and SEOS-V models. Comparing models with the same amount of $\Omega_{de}(a_0)$ we notice that SEOS-II increases the enhancement of the bump compared to  SEOS-I in table \ref{tab:seos}. The same happens when we compare model SEOS-IV with SEOS-III and model SEOS-VI with SEOS-V. To conclude, a steeper bump (larger $q$) shifts the peak to later times and increases the amount of extra energy density $\Omega_{ex}(a_c)$.

\subsection{Linear matter power spectrum} \label{linear-ps}

\subsubsection{Transitions at early times}

We clearly see  from   Fig.~\ref{fig:power-linear-bde} that a bump is indeed generated  corresponding to a wave-number of the order of $k_c$ 
where we plot  the ratio of the  matter power spectra $P_\text{BDE}/P_\text{$\Lambda$CDM}$.  
We plot the six BDE models in Table \ref{tab:bde}  with  solid blue for BDE-I; dashed red  for BDE-II; dash-dotted green for BDE-III; dotted magenta for BDE-IV; solid cyan  for BDE-V; and dashed yellow for BDE-VI.

We notice that for small scales, $k>k_c$, the deviation in the matter power spectrum between \LCDM and BDE is significant, peaking at mode $k_b$ which is of the order of $k_c$. However, well after the transition takes place,  corresponding to  modes $k\ll k_c$,  both models (\LCDM and BDE)  have the same $H$ and the ratio $P_\text{BDE}/P_\text{$\Lambda$CDM}\sim 1$.

The deviations at scales $k\geq k_c$ come from the dynamics of BDE and we can see the imprint left by the RDED  for modes  entering the horizon before $a_c$. The BDE models in Table \ref{tab:bde} peak  at $k_{b_L}=0.474  \,h \text{Mpc}^{-1}$ for models BDE I, II and III with a 
$k_c=0.148 \,h \text{Mpc}^{-1}$, while $k_{b_L}= 1.061 \,h \text{Mpc}^{-1}$ for models BDE IV,V and VI, with  and $k_c=0.295 \,h \text{Mpc}^{-1}$.   The enhancement is  correlated to the amount the diluted  component $\Oex$.
For instance, the case of BDE-I with $\Omega_{ex}^c=0.112$ has a bump increased at its peak by  $16\%$, while  BDE-II (BDE-III) have an enhancement of  $6.68\%$ ($2.5\%$), respectively. A similar pattern results for the BDE IV, V, and  VI models.

We have seen that the initial suppression of the linear evolution is reversed by the RDED and a bump is indeed generated at a linear level. The magnitude of the enhancement is related to the amount the diluted extra component $\Oex$  and the position of the peak of the bump given by $k_{b_L}$ is shifted to smaller scales, from $k_c=0.148 \iMpc $ to $k_{b_L}=0.917  \iMpc$ and from $k_c=0.295 \iMpc$ to $k_{b_L}=1.061 \iMpc $.

However, since these modes cross the horizon at early times,  they are no longer in the linear regime at present time. For this reason, in section \ref{oneloop} we use non-linear Standard Perturbation Theory (SPT) (see, e.g. \cite{Bernardeau_2002}) to analyse how much of this discrepancy is expected to be seen when we take into account the quasi-linear effects. This last effect is manifest in the matter power spectrum in figure \ref{fig:power-linear-bde}.

\begin{figure*}[ht]
        \includegraphics[height = 0.37\textheight]{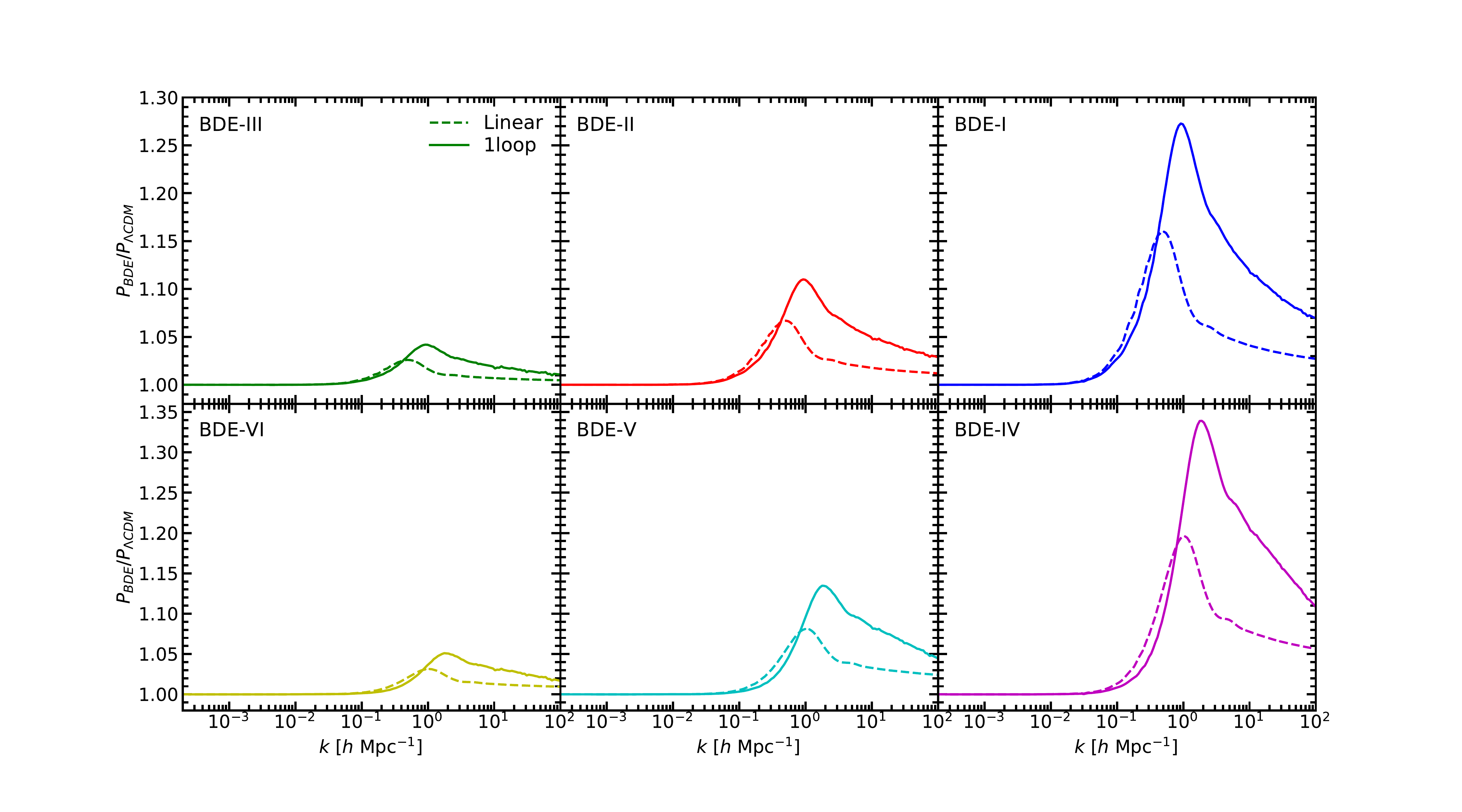}
\caption{\textbf{The 1-loop and linear power spectrum for BDE cosmologies.} We show the ratio $R(k)=P(k)_\text{BDE}/P(k)_\text{$\Lambda$CDM}$
at $z=0$ at a linear (dashed) and one-loop level (solid).  Top left: the BDE-III model; top middle: the BDE-II model; top right: the BDE-I model; bottom left: the BDE-VI model; bottom middle: the BDE-V model; bottom right: the BDE-IV model. Full curves are for 1-loop power spectrum; and dashed curves are for linear theory.}
\label{fig:power-1loop-linear-bde}
\end{figure*}

\subsubsection{Transitions at late times}\label{sec:latetimes}
We see that SEOS impacts the evolution of matter perturbations mainly at late times when the DE density is non-negligible. 
We take  SEOS as a model-independent smooth DE component, meaning that  $\delta \rho_{ex}=0$, since we are interested in parameterizing the dynamics of DE.
In appendix \ref{appendix} we present scalar field models which render the SEOS DE dynamics and the perturbations do not cluster.
We show in Fig.~\ref{fig:power-linear-seos}
the effects of SEOS  in the matter power spectra, where we display the ratio with respect to $\Lambda$CDM  for the SEOS models presented in table \ref{tab:seos}.
We work in all six models with the same amount of matter $\Omega_m h^2= 0.14$.

In the left panel
we show the ratio $P_{\text{SEOS}}(k)/P_{\text{$\Lambda$CDM}}(k)$ for different redshift values
($z = 0,\, 0.1,\, 0.2,\, 0.25,\text{and} , 5$) for SEOS-IV model. Blue curves
are for $z = 0$; red curves are for $z = 0.1$; green curves
are for $z = 0.2$; magenta curves are for $z = 0.25$; and
cyan curves are for $z = 5.0$. In the right panel we show
the same ratio for different models varying the quantity
of matter and rate of expansion (see Table \ref{tab:seos} IV A 1). Blue
curves are for SEOS-I; red curves are for SEOS-II; green
curves are for SEOS-III; magenta curves are for SEOSIV; cyan curves are for SEOS-V; and yellow curves are
for SEOS-VI.

At redshift $z=5$, during the Einstein-de Sitter phase, matter perturbations have almost the same  amplitude as in $\Lambda$CDM at all scales, the overall normalization is due to slight different rate of expansions.  

This overall suppression is due to faster expansion rate in \LCDM than in SEOS since the EoS
is always larger than for  cosmological constant , i.e $w_{SEOS} > -1$. We also notice that the overall suppression increases with time when DE starts to dominate. 
However, once the RDED in SEOS takes place at $a_c=0.8$ $(z_c=0.25)$ with a corresponding mode $k_c=a_cH(a_c)\simeq 3 \times 10^{-4} \iMpc$ it
generates bump in the ratio of power spectra with  a maximum at sligthly  smaller scales
at $k_b \simeq 5.5 \times 10^{-4} \iMpc$.

In the left panel of  fig.(\ref{fig:power-linear-seos}) we see snapshots  of the quotient
of the matter power spectrum at different values of $z$. Clearly the  bump 
is generated after the transition takes place at $z_c=0.25$, i.e for $z\leq z_c$. 
The evolution of the bump increases as $z \rightarrow 0$.
While  in the right panel of fig.(\ref{fig:power-linear-seos})  we see that the impact of the rapid dilution energy density appears for modes $k \geq  k_c$ entering the horizon slightly before the transition  occurs  at $a_c=0.8$.
 with  $k_c=a_cH(a_c)=3.27 \times 10^{-4}\iMpc=2.22 \times 10^{-4}$ Mpc$^{-1}$ for SEOS-IV model.

Notice that the amount of matter $\rho_m(a_0)$ is the same in all models in  table \ref{tab:seos}  and it is the dynamics of the DE  component what decreases the amplitude by the same amount  for all modes, as compared to a \LCDM model.

However, the  impact on the evolution of the matter perturbations due to the RDED generates the bump seen at $k\sim 5 \iMpc$.
In fact, we can isolate the effects of the background expansion and growth of fluctuations by looking at the spectra at late times, when DE is dominant. For instance, at $z=0.25$ the spectrum in SEOS is suppressed by $ ~1.6\% $ on all scales due to the late-time dynamics of DE, but this suppression is overwhelmed on large scales by the enhancement effect due to SEOS rapid dilution at $z_c=0.25$, which lead to a small excess of power at $k_b=6.049 \times 10^{-4}\iMpc$. At smaller redshifts, the spectra continue to decline overall, but increasing at the bump location.

Finally it is worth remarking that SEOS and  $\Lambda$CDM  share the same set of cosmological parameters  $h, w_m=\Omega_m h^2$ at present time, however the evolution
differs due to the EoS in eq.(\ref{wseos}).  Higher values of $q$  enhanced slightly  the amplitude  of bump and increases the value of $k_{b,L}$. Finally we remark  that $k_{b,L}$ is of the same order as $k_c$ but  slightly shifted to smaller scales, as described in the analytic presentation in section \ref{sec.P}. 

\begin{figure*}[ht]
        \includegraphics[height = 0.37\textheight]{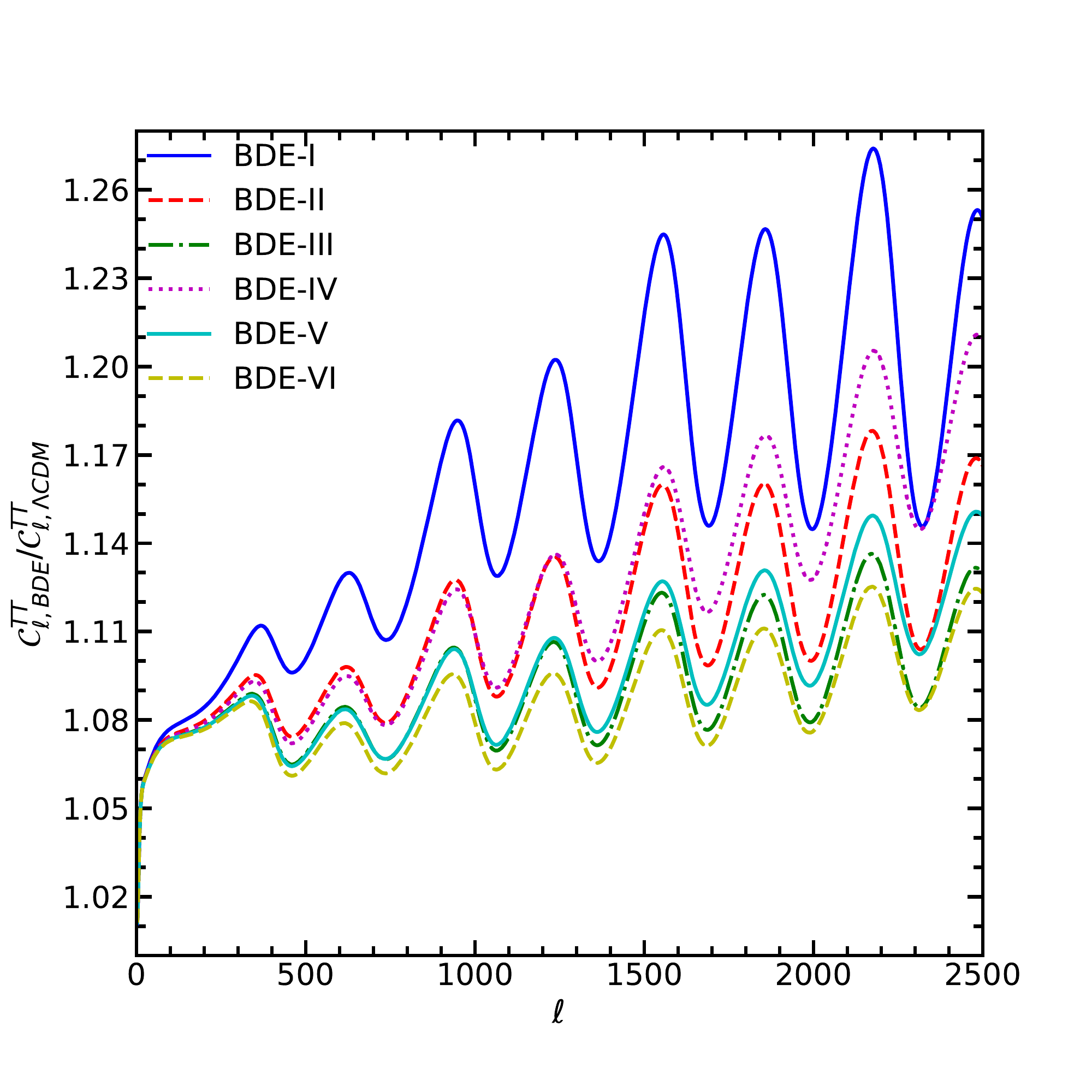}
\caption{\textbf{The cosmic microwave background spectra for BDE cosmologies.} The ratio $\mathcal{C}_{l, \text{BDE}}^{TT}/\mathcal{C}_{l, \Lambda\text{CDM}}$. Solid blue curves are for BDE-I; dashed red curves are for BDE-II; dashed-dotted green curves are for BDE-III; dotted magenta curves are for BDE-IV; solid cyan are for BDE V; and dashed yellow are for BDE-VI. }
\label{fig:cmb-bde}
\end{figure*}

\begin{figure*}[ht]
        \includegraphics[height = 0.37\textheight]{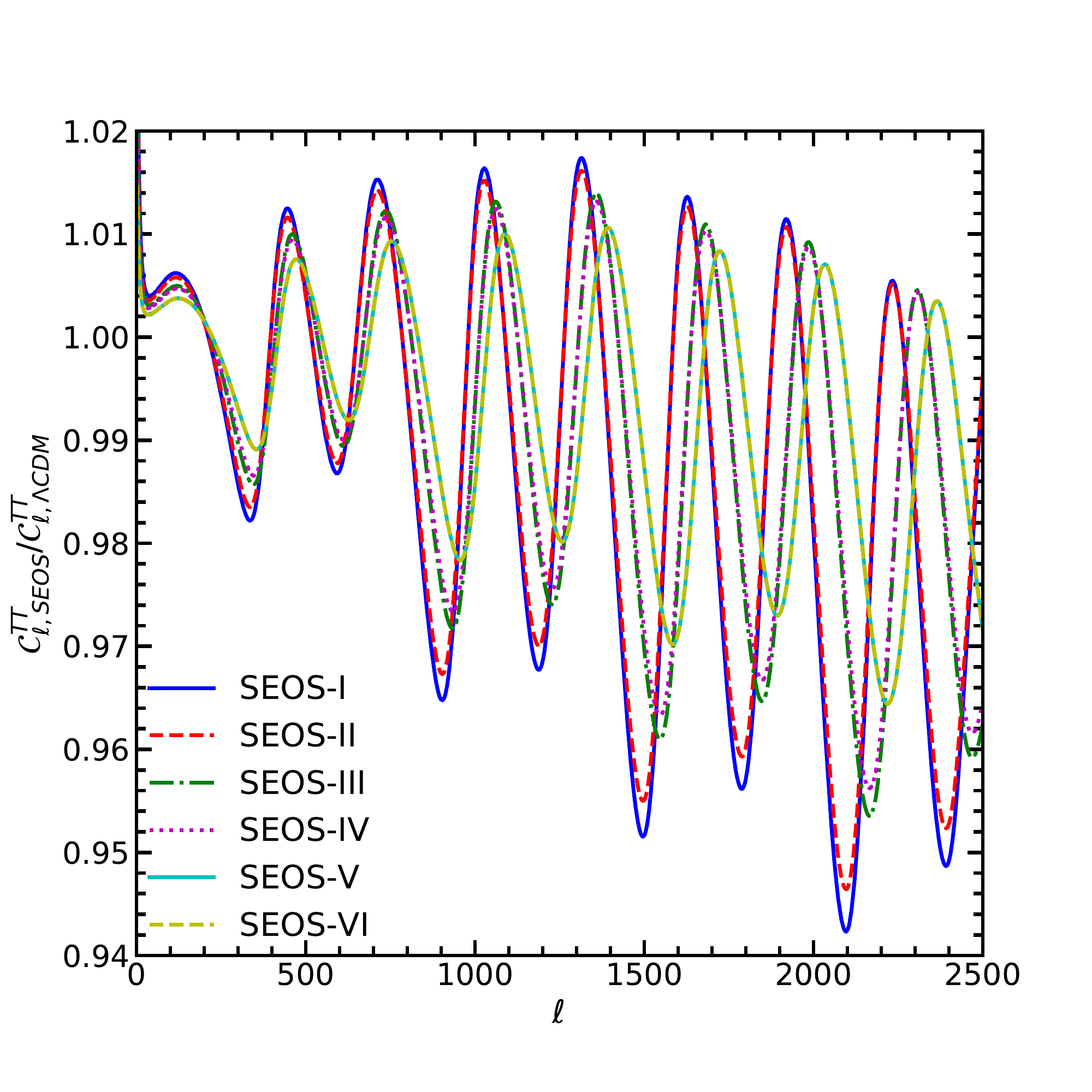}
\caption{\textbf{The cosmic microwave background spectra for SEOS cosmologies.} The ratio $\mathcal{C}_{l, SEOS}^{TT}/\mathcal{C}_{l, \Lambda CDM}$. Solid blue curves are for SEOS-I; dashed red curves are for SEOS-II; dashed-dotted green curves are for SEOS-III; dotted magenta curves are for SEOS-IV. } 
\label{fig:cmb-seos}
\end{figure*}

\subsection{Nonlinear evolution in SPT}\label{oneloop}


%
In this section we compute the non-linearities in the power spectrum for BDE models using 1-loop SPT. For this we use the publicly available code \texttt{MGPT}\footnote{\href{https://github.com/cosmoinin/MGPT}{https://github.com/cosmoinin/MGPT}} \cite{Aviles:2017aor, Aviles:2018saf} that accounts for a background evolution different than $\Lambda$CDM. 

We work out only the BDE models since these have the bumps located at quasi-linear and non-linear scales. In turn, the SEOS the transition modes enter at very late times and henceforth the signatures are located at very large scales and are not affected by non-linearities.  
In Fig.~\ref{fig:power-1loop-linear-bde} we show the ratios  $R(k)=P(k)_\text{BDE}/P(k)_\text{$\Lambda$CDM}$ for the different models. 
Solid curves are for 1-loop power spectrum and dashed curves are for linear theory. We show that the perturbative 1-loop affects the intermediate scales and small scales. 

The 1-loop power spectrum of BDE-I, BDE-II, and BDE-III has a bump contributing $27.2\%$, $11.0\%$, and $4.10\%$ to the power spectrum at $k=0.917\iMpc$, respectively. A similar pattern of results was obtained for the BDE-IV, BDE-V, and BDE VI models whose contribute $33.9\%$, $13.4\%$, and $5.10\%$ to the power at $1.908\iMpc$, $1.773\iMpc$, and $1.773\iMpc$, respectively. We notice that nonlinearities tend to shift the peak of the bumps to smaller scales and enhance their amplitudes with respect to the linear theory.

\subsection{CMB power spectrum}\label{cmb-ps}

We present the effect of  having a RDED in the CMB power spectra in BDE ads SEOS cosmologies.

\subsubsection{Transitions at early times}

We show  the temperature power spectrum of the CMB for different BDE cosmologies in Fig. \ref{fig:cmb-bde}, where we plot the ratio $C^{TT}_{\ell,BDE}/C^{TT}_{\Lambda CDM}$. An increase in the amount of radiation at early times, before recombination, can be seen to affect the CMB: an angular shift to higher $\ell$s, an enhancement in the amplitude,  and a change in the damping scale.

Having an extra radiation term in this model shifts the CMB peaks. It is well known \cite{Hu:1995fqa} that the angular position of the oscillations peaks are located at the extrema of the oscillations given by $k r_s(a_*)$, where $r_s(\ad)$ is the sound horizon at last scattering, given in Eq.~\eqref{rs}, which depends on the expansion rate $H$. Because we increase the radiation content before the last scattering, for $a < a_c < a_*$, the expansion rate is increased (see Eq.~\eqref{aa}) making the scattering surface thinner in BDE cosmologies than in $\Lambda$CDM. Thus, slightly shifting the peaks of the oscillation to higher $\ell$s.

The amplitude of the CMB peaks depends on the matter-radiation ratio, therefore it relies on the scale parameter $a_c$, and $\Oex$. A change in the diffusion damping scale, given by
\be\label{diffusion}
r_d^2= (2\pi)^2\int_0^{\ad} \fr{da}{a^3H n_e \sigma_T}\le(\fr{R^2+ \fr{16}{15} (1+R)}{6(1+R^2)}
\ri)\ee
with $R = (3\rho_b)/(4\rho_\gamma)$, $n_e$ the number density of free electron and $\sigma_T$ the Thompson cross-section, is also noticeable, where all oscillation modes are larger than in $\Lambda$CDM.  
The damping factor at last scattering depends on two things, first the visibility function that remains independent of cosmological parameters before the transition $a_c$, and second, the behavior of the damping scale $k_D$ through last scattering. The last one depends on the acoustic scale and, on a smaller degree, on the Hubble rate \cite{Hu:1995fqa}. Since the extra radiation term $\rex$ changes both of these parameters, it is expected to have a smaller damping factor than in  $\Lambda$CDM.

The overall change can be noticed in Fig. \ref{fig:cmb-bde}. In particular, for the mode at $k=0.148$h $M_{pc}^{-1}$ we obtain an increment in power compared to \LCDM. The increment in power becomes larger with increasing $\ell$. For example, taking the  multipole  $\ell\sim 2000$  we find on average an enhancement  close to 20\%, 14\% and 10\%  for  models BDE-I, BDE-II, and BDE-III, respectively, compared to \LCDM. In turn, the increase for the mode $k=0.295hMpc^{-1}$ is slightly smaller  with 17\%, 13\%, and 10\%  for BDE-IV, BDE-V, and BDE-VI models, respectively.

\subsubsection{Transitions at late times}

Here, we show how the dark energy parametrization, SEOS, affects the CMB TT anisotropies. In Fig.~\ref{fig:cmb-seos} we show $\mathcal{C}^{TT}_{l,SEOS}/\mathcal{C}^{TT}_{l,\Lambda CDM}$ for different combinations of parameters in the SEOS model, varying the quantity of matter and rate of expansion.

We see that SEOS impacts the temperature field not only by changing the amount of matter, but there is also a distinctive imprint left by the steepness parameter $q$. Similar as in the BDE model, see sub-section \ref{cmb-ps}, the SEOS model introduces the same three characteristics in the CMB due to $\rex$. First, the shift to larger $\ell$s of the wiggles due to the change of the expansion rate caused by the increase of  radiation content before the last scattering, and therefore, changing the extrema position of each oscillation as seen in Fig.~\ref{fig:cmb-seos}. 
Larger values of the steepness $q$ of the dark energy parametrization decreases the amplitude of the oscillations as we can see in the solid blue and dashed red curves, while smaller amount of DE also reduces the amplitude. It is worth keeping in mind that all six SEOS models have the same amount of matter $\Omega_m h^2=0.141$.

\section{Conclusions}\label{sec.conclusion}

In this work, we studied cosmological implications of introducing an extra energy density $\rex$ to a $\Lambda$CDM model that dilutes rapidly  at given scale factor $a_c$ and corresponding wave-vector amplitude $k_c=a_cH(a_c)$.
This RDED transition affects the background evolution, and hence the cosmological distances, such as the comoving angular distance $D_A$, the angular diameter $D_M$, the acoustic scale at recombination $r_s$ and the diffusion damping scale $r_d$.  Furthermore, the rapid dilution of $\rex$ also impacts the evolution of the matter perturbations and CMB power spectrum. This RDED leaves distinctive  features in the matter power spectrum. In particular, it generates a 
bump in the matter power spectrum, more visible once we compare to $\Lambda$CDM. 
The bump is generated because the linear growth of $\delta_m(k)$ has a higher increase rate for larger $H$ in radiation domination
(c.f. Eq.~(\ref{deltatau})). 
This increase takes place only for modes $k\geq k_c $ when $\rex >0$ while for mode $k<k_c$  both models have the same  expansion rate and the evolution of $\delta_m(k)$ is the same for both cases.  The amplitude of the bump is related to the amount of the diluted energy density $\rex$, while the mode is located about $k\sim k_c$. We study these bumps in the linear regime and also apply one-loop corrections using SPT.

We concentrated here in two different models, one located at large scales and the others at small scales.
Both cases are inspired by the BDE model  \cite{Almaraz:2018fhb,delaMacorra:2018zbk}, where an RDED phase transition takes place in the radiation dominated epoch with $a_c\sim 10^{-6}$, affecting modes entering the horizon at early times. The corresponding mode is $k_c\sim 1 \iMpc$ and a bump is generated for modes $k\geq k_c$. Interestingly the BDE model also shows a RDED  with a steep transition centered at a redshift  $z_c \approx 0.625$ resulting in second bump located at large scales. 

To study the imprints of dark energy, we considered a model-independent analysis and we parameterized the dynamics of dark energy using a
steep equation of state (SEOS) \ci{Jaber:2017bpx}. In the SEOS model the  EoS  $w$ has a transition at  $z_c=0.28$. Here we studied two extreme
cases, with soft and steep transitions, to observe different impacts on the CMB and matter power spectra.

We computed the position and amplitude of the peaks of the bumps in the matter power spectrum through the one-loop correction analysis using the code \texttt{MGPT}, taking into account the background evolutions with and without $\rex$. We conclude that the non-linear effects shift the peak of the bumps to smaller modes and enhance their amplitudes in comparison to linear computations.

We took different values of cosmological parameters for our two models shown in Table \ref{tab:bde} for small scales, and  \ref{tab:seos} for large scales. In the linear regime, we  see the bump imprint left by the rapid dilution of BDE on modes $k > k_c$, entering the horizon before $a_c$, as a peaked  bump centered at $k = 0.47 \,h \text{Mpc}^{-1}$ for BDE-I,BDE-II,BDE-III ($k = 1.06 \,h \text{Mpc}^{-1}$ for BDE-IV,BDE-V, BDE-VI), where the enhancement of power in BDE is about $16\%$ ($19.58\%$), respectively. We see deviations at larger scales e.g. $k \geq  0.05 \,h \text{Mpc}^{-1}$  and $k \geq  0.5 \,h \text{Mpc}^{-1}$  respectively, due to the width of the bump in each case. The scale of the bump is mainly located at non-linear scales, although covering also quasi-linear scales, and once we determine the one-loop power spectra we find that BDE-III (BDE-VI) provides an enhancement of $4\%$ ($5\%$) at $k=0.9\iMpc$ ($1.8\iMpc$), respectively. Nevertheless, these results should be taken as indicative since the high-$k$ tails of the bumps are out of the reach of perturbation theory. 

The late time dark energy transitions impacts the evolution of matter perturbations mainly at late times during the Einstein-de Sitter phase, matter perturbations have the same amplitude as in $\Lambda$CDM at all scales, the overall normalization is due to a slightly different rate of expansion in  $\Lambda$CDM  and  SEOS models.  Here we studied the same mode $k= 8.37 \times 10^{-5} \,h \text{Mpc}^{-1}$ but different amounts of matter and steepness of the transition.

As time increases,  the effect of the dynamical dark energy of the SEOS model is altered decreasing the amplitude for all Fourier modes and originating a bump at $k \approx 5 \times 10^{-3}$; that is, the power spectrum of $\Lambda$CDM is always larger than the power spectrum of SEOS model originating a bump at lower redshift.

Summarizing, this distinctive signature, named as bump, has been studied at a linear and one-loop level in perturbation theory. We leave for a future work the study of bump cosmologies with the use of halo-based models in order to probe the non-linear regime \cite{Dante2020}, as well as its impact towards a possible solution to the $H_0$ crisis \cite{h0tensionAx}.  To conclude, a small amount of extra energy density that dilutes rapidly is consistent with present-day cosmological measurements and may solve tensions in cosmology leaving distinctive detectable signatures.

\section*{Acknowledgements}

AM and DG acknowledge partial support from Project IN103518 PAPIIT--UNAM and AM  from PASPA--DGAPA, UNAM and CONACyT.
AA acknowledges partial support from Conacyt Grant No. 283151
MJ acknowledges the support of the Polish Ministry of Science and Higher Education MNiSW grant DIR/WK/2018/12. 
Part of this work was supported by the ``A next-generation worldwide quantum sensor network with optical atomic clocks'' project, which is carried out within the TEAM IV programme of the Foundation for Polish Science co-financed by the European Union under the European Regional Development Fund. DG thanks support from a CONACyT PhD fellowship.
JM acknowledges Catedras-CONACYT financial support and MCTP/UNACH as the hosting institution of the Catedras program.
EA thanks CONACyT for a Postdoctoral scholarship.

\section{Appendix}\label{appendix}

\subsection{ Rapid Diluted Energy Density (RDED):\\ a Scalar Field approach }

There are several scenarios on how one can  generate a rapid dilution of an energy density $\rex$  with  $\Oex(a_f) \ll \Oex(a_i)$
 where this dilution is due to a change in the EoS $w$ due to the underlying dynamics of the this fluid with  $\Delta w=w_f-w_i  >0 $.
Here we will work with a scalar field  $\phi$, with energy density   $\rho_\phi =E_k + V(\phi)$ and pressure $p_\phi= E_k - V(\phi)$ giving  an EoS $w=p_\phi/\rho_\phi$. 
Depending on the choice of scalar potential $V(\phi)$ and kinetic term $E_k$ the
 dynamical evolution of the scalar field $\phi$ has  been extensively studied in the literature in the last 20 years \ci{Zlatev:1998tr,Steinhardt:1999nw,delaMacorra:1999ff}, 
 \ci{delaMacorra:2001xx, delaMacorra:2004mp,DelaMacorra:2001uq} and  the review work\ci{Copeland:2006wr}.

Setting $\phi (t,x)= \phi (t) +\delta \phi (t,x)$,  the evolution of the homogeneous scalar field $\phi (t)$  is given by the Klein Gordon equation
\begin{equation} \label{eq:klein_gordon} 
\ddot{\phi} + 3H\dot{\phi}+\tfrac{dV}{d\phi} = 0,
\end{equation}
where we used a canonical kinetic terms  one has $E_k = \tfrac{1}{2}\dot{\phi}^2$ with 
\begin{equation} \label{eq:friedmann_after} 
H^2 = \tfrac{8\pi G}{3}(\rho_{mo}a^{-3}+\rho_{ro}a^{-4}+ \rho_\phi), 
\end{equation}
and  $\rho_{m}$ and $\rho_{r}$ corresponds to matter  and radiation  contribution. 
Widely used scalar potentials  are either exponential  or inverse power terms,    e.g.
 \be
 V(\phi)=\Lambda^4  e^{-\alpha \phi}, \;\;\;\;\   V=\Lambda^{4+n}  \phi^{-n}
 \ee
with $\alpha, n$ dimensionless constants while  $\Lambda$ having mass dimension. This potentials have different behaviours, where they can track the background
(either radiation or matter with $w_i=0$ or $w_i =1/3 $,    and can later leap to $w_f=1$  (i.e. $V\gg E_k$) for a long period of time and diluting $\rho_\phi$,
generating a rapid dilution $\rex$  situation.

As an interesting example we  present the Bound Dark Energy (BDE) model\ci{Almaraz:2018fhb,delaMacorra:2018zbk} where an extra gauge Dark Group (DG) is proposed which contains  massless particles weakly coupled at high energies  and  contribute to the total content of radiation of the Universe  but at lower energies the strength of the DG interaction increases and the  gauge coupling  becomes strong at the condensation energy scale $\Lc$  and scale factor  $a_c$.   At this scale   the fundamental particles form gauge invariant  composite states,  dark mesons and dark baryons,  which  acquire a non-perturbative mass  proportional to $\Lc$ as for example for protons, neutrons and pions in QCD. This is  expected  from dimensional analysis of spontaneous symmetry breaking theories were we  expect the relevant physical quantities to be  proportional to the symmetry breaking scale. 
The energy stored in DG is transferred to the lightest composite field,  corresponding to a pseudo Goldstone boson field, our BDE scalar  $\phi$. Below the condensation scale  (i.e. for $a\geq \ac$) we describe the evolution of the scalar field $\phi$   by the Klein Gordon equation with an inverse power law potential  (IPL)
$ V(\phi)=\Lc^{4+n}\;\phi^{n}$ with $n=2/3$. 
The initial conditions at  $a_c$ are $\phi_c(\ac)=\Lc$, $\rho_{\phi}(a_c)=2\Lambda_c^4/(1-w_{\phi c})=3\Lc^4$ and  $\dot{\phi}(a_c)=\sqrt{2\Lambda_c^4(1+w_{\phi c})/(1-w_{\phi c)}}=2\Lc^2$,
and $w_{\phi c}=1/3$. 
The solution to the dynamical evolution has en $w_\phi =1/3$ for $a<a_c$, i.e. before the transition takes place at $a_c$. Once the IPL potential $V(\phi)$  at $a_c$
is dynamically generated the  EoS $w_\phi\equiv p_\phi/\rho_\phi$ leaps to $w_\phi =1$  and remains at this value for a long period of time. 
The energy density  $\rho_\phi$ quickly dilutes as  $\rho_\phi\propto a^{-6}$  due to the change in the  EoS  from $w_\phi(a<a_c)=1/3$ to $w_\phi=1$ 
lasting a  long period of time. 
In this example the transition takes place in radiation domination era so we have $w_i=1/3$ and a final  $w_f=1$ and $\Delta w= w_f-wi=2/3  > 0 $ and we have a rapid diluted 
energy density $\rho_\phi$ with $ \Oex (a_c)\simeq 0.1$ to $\Oex(a_f) \simeq  0$.  The steepness of this transition is set by the order parameter 
$\lambda (\phi)= - m_{pl}(dV/d\phi)/V$  at $a_c$, and for IPL potentials we have $\lambda (\phi)= n (m_{pl}/\phi)$ .with $\lambda (\phi_c)\gg 1$ since $\phi_c\ll m_{p|}$.
This dilution has an impact on distances and generates a bump in the power spectrum at $k_c\equiv a_cH(\ac)$Mpc$^{-1}$.

At a later stage  the EoS $w_\phi$  in IPL potentials drops to  $w_\phi=-1$ (the potential energy $V$ dominates over the kinetic term $E_k$) and $w_\phi$ remains at this valid  for long period of time. 
Finally, close to present time the EoS grows from $w_\phi =-1$ to $w_{\phi o} \simeq  - 0.9$, giving a $\Delta w =w_f-w_i=0.1 >0$) close to present times with  $a_i=2/3, a_f=1$)
corresponding to a small $k\sim 10^{-4}$ Mpc$^{-1}$.  This second rapid dilution $\rex$ has a much smaller effect in the power spectrum since the amount of $\Oex$ is small but has an important impact on distances and affect for example BAO measurements.
BDE corresponds to a specific model  where we have relativistic particles with $\Omega(a_c)=0.112$ for $a<a_c$ and phase transition takes place at $a_c$  due to the underlying dynamics of the BDE model. A  scalar potential is dynamically  formed  with a derived expression $V(\phi)=\Lambda_c^{4+2/3}\phi^{-2/3}$, with $\Lambda_c = 44.09\pm 0.28 eV$,  $a_c=(2.48\pm 0.02) 10^{-6}$  and a mode $k_c=0.925$ hMpc$^{-1}$ \cite{Almaraz:2018fhb,delaMacorra:2018zbk}. 

The  Bound Dark Energy  we just  present is an example of a DE model derived from particle physics which contains   two regions where  we naturally obtain a rapid dilution of $\rho_\phi$ with a
 $\Delta w >0$  and therefore we have two signatures in the power spectrum (at very different scales $k$)  and  also have  an interesting impact  on cosmological distances
 in the region $0< z< 1$  allow for a better cosmological fit than $\Lambda$CDM model.

\subsubsection{ Perturbations for Scalar Fields }

We stay in the linear regime where the energy density and other quantities can be decomposed into a homogeneous part (commonly denoted with a bar) and a small position-dependent perturbation. We solve the perturbed equations in the syncrhonous gauge defined by the line element \cite{Ma:1995ey}
\begin{equation} \label{eq:syn_gauge}  ds^2=a^2(\tau)(-d\tau^2+(\delta_{ij}+h_{ij})dx^idx^j),\end{equation}
where $d\tau\equiv dt/a$ denotes the conformal time. We assume initial adiabatic conditions and make the ansatz that before $a_c$ the DG perturbations mimic the perturbations of neutrinos. Once the condensation happens, the perturbations in the energy density and pressure of the scalar field are given by:
\begin{equation} \label{eq:delta_phi} \delta\rho_\phi = \frac{\bar{\phi}'\delta \phi'}{a^2}+\frac{dV}{d\phi}\delta \phi, \hspace{0.5cm} \delta P_\phi = \frac{\bar{\phi}'\delta \phi'}{a^2}-\frac{dV}{d\phi}\delta \phi, \end{equation}
where the primes stand for conformal time derivatives. The evolution of $\delta \phi$ in Fourier space is determined by
\begin{equation} \label{eq:kg_perturbed} \delta \phi'' + 2\mathcal{H}\delta \phi' + (k^2+a^2\tfrac{d^2V}{d^2\phi})\delta \phi = -\tfrac{1}{2}\bar{\phi}'h' \end{equation}
Here $\mathcal{H} \equiv a'/a$ is the conformal expansion rate and $h=Tr(h_{ij})$. The term proportional to $k^2$
in eq.(\ref{eq:kg_perturbed}) constrains the formation of structure formation which in DE parametrization, such as in CPL
$w=w_o+w_a(1-a)$ or SEOS, would blow up when $w<0$. Therefore scalar fields yield viable DE models as shown in BDE where the DE perturbations do contribute but at a small percentage level \cite{Almaraz:2018fhb,Almaraz:2019zxy}.
The scalar field enters the perturbation equations of the other fluids \cite{Ma:1995ey} via $\mathcal{H}$ and through extra source terms proportional to $\delta \phi$ and $\delta \phi'$. For example, the CDM overdensities $\delta_c$ evolve according to
\begin{equation} \label{eq:CDM_perturbed} \delta_c'' + \mathcal{H}\delta_c'-\tfrac{3}{2}\mathcal{H}^2\sum\Omega_i\delta_i (3c_{s,i}^2+1)=0, \end{equation}
where the sum runs over all the fluids with sound speed $c_{s,i}^2=\delta P_i/\delta \rho_i$.

\bibliography{bump}

\end{document}